\documentstyle[12pt]{report}
\input tcilatex

\begin{document}

\date{}

\centerline{\LARGE \bf Locally Anisotropic Interactions:} \vskip4pt
\centerline{\LARGE \bf III. Higher Order Anisotropic} \vskip4pt
\centerline{\LARGE \bf Supergravity} \vskip10pt
\centerline{\sf Sergiu I. Vacaru }

\vskip8pt
\centerline{\noindent{\em Institute of Applied Physics, Academy of Sciences,}}
\centerline{\noindent{\em 5 Academy str., Chi\c sin\v au 2028,
Republic of Moldova}} \vskip8pt
\centerline{\noindent{ Fax: 011-3732-738149, E-mail: lises@cc.acad.md}}
\vskip10pt {\footnotesize {\bf Abstract.} A general approach to formulation
of supergravity in higher order anisotropic superspaces (containing as
particular cases different supersymmetric extensions and prolongations of
Riemann, Finsler, Lagrange and Kaluza--Klein spaces) is given. We analyze
three models of locally anisotropic supergravity.}

\section*{1. Introduction}

This is the third of a series of papers that attempt to develop the basis
for formulation of physical theories with supersymmetric locally anisotropic
interactions. The first paper \cite{vjph1} in the series is devoted to
definition of higher order anisotropic superspaces and to investigation of
geometric properties and possible physical significance of nonlinear
connection (in brief, N--connection) superfield (s--field) in such
superspaces (s--spaces). Here we note that higher order anisotropies are
modeled on higher order tangent superbundles or on corresponding
generalization of vector superbundles (vs--bundles), called distinguished
vector superbundles (dvs--bundles), provided with compatible N--connection,
distinguished by the N--connection linear connection and metric structures.
The aim of the second paper \cite{vjph2} is to continue our investigations
of basic geometric structures on higher dimension superspaces with generic
local anisotropy and to explore possible applications in theoretical and
mathematical physics (see our works \cite{vjmp,vg,vlasg,voa,vsp96,vcl96}).
We have computed torsions, curvatures in dvs--bundles and defined components
of Bianchi identities and Cartan structure equations.

In this paper we consider the problem of construction of higher order
an\-isot\-rop\-ic supergavity. Our main idea is to use the N--connection as
a s--field splitting ''step by step'' the higher dimensional s--space into
lower dimensional ones. Introducing corresponding parametrizations of the
N--connection components we can modelate, as particular cases, different
variants of higher dimensional compactification and reduction to physical
space--times. We shall develop a general supersymmetric differential
geometric techniques for investigation of locally anisotropic field
interactions in dvs--bundles. A great part of our considerations will be
devoted to the higher order anisotropic generalization of {\sf N}=1
s--gravity (for simplicity, we shall use as a starting point the
Dezer--Zumino model \cite{des} in the framework of, supersymmetric in our
case, osculator bundles \cite{ma87,ma94,mirata}). In order to have the
possibility to compare our model with usual {\sf N}=1 ( one dimensional
supersymmetric extensions; see, for instance, \cite{wes,mul,sal})
supergravitational models we develop a supergavity theory on osculator
s--bundle $Osc^z\widetilde{M}_{(M)}$ \cite{vjph1,vjph2}, where the even part
of s--manifold $\widetilde{M}_{(M)}$ has a local structure of Minkowski
space with action of Poincare group. In this case we do not have problems
connected with definition of spinors (Lorentz, Weyl or Maiorana type) for
spaces of arbitrary dimensions and can solve Bianchi identities. As a matter
of principle, by using our results on higher dimensional and locally
anisotropic spaces \cite{vjmp}, we can introduce distinguished spinor
structures and develop variants of extended supergravity with general higher
order anisotropy. This approach is based on global geometric constructions
and allows us to avoid tedious variational calculations and define the basic
field equations and conservation laws on s--spaces with local anisotropy. We
shall present a geometric background for extended supersymmetric locally
anisotropic models.

We shall begin our study with N--connection s--spaces in section 2. Such
s--spaces are generalizations of flat s--spaces containing a nontrivial
N--connection structure but with trivial d--connection. We shall introduce
locally adapted s--vielbeins and define s--fields and differential forms in
N-connection s--spaces. Sections 3 and 4 will by correspondingly devoted to
gauge s-field and s--gravity theory in osculator s--bundles. Bianchi
identities and constraints in osculator s--gravity will be solved in section
5. In order to generalize our considerations for higher order supersymmetry,
in section 6, we shall introduce Einstein--Cartan equations on distinguished
vector superbundles (locally parametrized by arbitrary both type commuting
and anticommuting coordinates) in a geometric manner, in some line following
the geometric background for Einstein relativity, but in our case on
dvs-bundles provided with arbitrary N--connection and distinguished torsion
and metric structures. We can consider different models, for instance, with
prescribed N--connection and torsions, to develop a Einstein--Cartan like
theory, or to follow approaches from gauge gravity. In section 7 we shall
propose a variant of gauge like higher anisotropic supergravity being a
generalization to dvs--bundles of models of locally anisotropic gauge
gravity \cite{vg}.

\section*{2. N--connection Superspaces}

The reader shall consult basic definitions and denotations on distinguished
vector and osculator s--bundles from \cite{vjph1,vjph2}. In subsection 2.1
we shall outline some necessary formulas and definitions, the rest of the
section is an introduction into the geometry of N--connection s--bundles.

\subsection*{2.1 Geometric objects in dvs- and osculator s--bundles}

Coordinates on a dvs--bundle $\widetilde{{\cal E}}^{<z>}$ are denoted
$$
u^{<\alpha >}=(u^{\alpha _p},y^{A_{p+1}},...,y^{A_z})=(x^I,y^{<A>})=
$$
$$
(\widehat{x}^i,\theta ^{\widehat{i}},\widehat{y}^{a_1},\zeta ^{\widehat{a}%
_1},...,\widehat{y}^{a_p},\zeta ^{\widehat{a}_p},...,\widehat{y}^{a_z},\zeta
^{\widehat{a}_z}),
$$
where indices $I,A_{1,}A_{2,...,}A_p,...,A_z$ run from 1 to the dimension of
superspaces used for construction of s--bundles into consideration and $%
<\alpha >$ and $<A>$ are cumulative indices, subindex $p$ enumerates the
number of anisotropic ''shells'' of higher order anisotropic s--spaces and
takes values from 1 to an integer $z$ (we shall also write p=0 for indices
on the base s--space, i.e. $A_0=I,$\thinspace and note that according to
conventions from our previous works \cite{vlasg,vjph1} we use Greek indices
without brackets for geometrical objects on vs--bundles).

A N--connection is given locally by a matrix $N_{<\alpha >}^{<\beta >}$ with
components $N_{A_f}^{A_p},p>f,$ or by the inverse N--connection matrix $%
M_{A_p}^{A_f},$ and defines respectively the so--called locally adapted
bases and dual bases (in brief, la--frame, or la-base)%
$$
\delta _{<\alpha >}=N_{<\alpha >}^{<\beta >}\partial _{<\beta >}=\{\partial
_{A_f}-N_{A_f}^{A_p}\partial _{A_p},p>f\}\eqno(1)
$$
and
$$
\delta ^{<\alpha >}=M_{<\beta >}^{<\alpha >}d^{<\beta
>}=\{d^{A_f}+M_{A_p}^{A_f}d^{A_p},f>p\}\eqno(2)
$$
where $\partial _{<\beta >}=\frac \partial {\partial u^{<\alpha >}}$ and $%
d^{<\beta >}=du^{<\beta >}$ are usual local coordinate partial derivation
and differential operators.

The unhlolonomic coefficients $w_{<\alpha ><\beta >}^{<\gamma >}\,$ are
introduced by using the supersymmetric anticommutation of bases (1)%
$$
[\delta _{<\alpha >},\delta _{<\beta >}\}=\delta _{<\alpha >}\delta _{<\beta
>}-(-)^{|<\alpha ><\beta >|}\delta _{<\alpha >}\delta _{<\beta >}=w_{<\alpha
><\beta >}^{<\gamma >}\delta _{<\gamma >},\eqno(3)
$$
where we write, for simplicity, $(-)^{|<\alpha ><\beta >|}=(-1)^{|<\alpha
>||<\beta >|},|<\alpha >|=0$ for even values of $<\alpha >$ and $|<\alpha
>|=1$ for odd values of $<\alpha >.$

A distinguished connection $D=\{D_{<\alpha >}\}$ (d--connection) is a linear
connection adapted to N--connection structure; the d--connection
coefficients are defined with respect to a la--base (1):%
$$
D_{\delta _{<\beta >}}\delta _{<\gamma >}=\Gamma _{<\gamma ><\beta
>}^{<\alpha >}\delta _{<\alpha >}.\eqno(4)
$$

A distinguished metric $G$ (d--metric) is represented as
$$
G=g_{<\alpha ><\beta >}\delta ^{<\alpha >}\otimes \delta ^{<\beta >}.\eqno(5)
$$
We shall also use the inverse metric $g^{<\alpha ><\beta >}.$

D--connection and d--metric structures are compatible if there are satisfied
conditions%
$$
D_{<\alpha >}g_{<\gamma ><\beta >}=0.
$$

The torsion $T_{<\beta ><\gamma >}^{<\alpha >}$ and curvature $R_{<\beta
><\gamma ><\delta >}^{<\alpha >}$ coefficients for a d--connec\-ti\-on $%
\Gamma _{<\gamma ><\beta >}^{<\alpha >}$ have been computed in explicit form
in our paper \cite{vjph2} ( formulas (10) and (14)). In brief, we shall use
instead of distinguished super--(vectors, geometrical objects)tensors on
dvs--bundles terms ds--(vectors, geometrical objects)tensors and so on.

Roughly speaking, osculator bundles $Osc^{<z>}\widetilde{M}$\ are usual
dvs--bundles for which the dimensions of higher order fibers coincides with
the dimension of base s--manifold (see the rigorous definition and detailes
in \cite{vjph1,vjph2}).

\subsection*{2.2 Supervielbeins in N--connection s--spaces}

Before we continue our analysis of the locally anisotropic superspaces, we
shall first consider dvs--spaces $\widetilde{{\cal E}}_N$ provided with
N--connection structure, having trivial (flat like with respect to
la--frames (1)) d--connection (4) and constant d--metric (5). Such spaces
will be called N--connection superspaces and denoted as $\widetilde{{\cal E}}%
_N^{<z>}.$

We shall use la--frame decompositions of metric on $\widetilde{{\cal E}}%
_N^{<z>}:$%
$$
G_{<\alpha ><\beta >}(u)=l_{<\alpha >}^{<\underline{\alpha }>}(u)l_{<\beta
>}^{<\underline{\beta }>}(u)G_{<\underline{\alpha }><\underline{\beta }>}.%
\eqno(6) 
$$
Indices of type $<\underline{\alpha }>,<\underline{\beta }>,...$ are
considered as abstract ones \cite{pen,penr1,penr2} s--vielbein indices. On
N-connection s--spaces we can fix la--frames and s--coordinates with
constant s--vielbein components $l_{<\alpha >}^{<\underline{\alpha }>}$ and
d--metrics $G_{<\underline{\alpha }><\underline{\beta }>}$ and $G_{<\alpha
><\beta >}.$

We suppose that s--space $\widetilde{{\cal E}}_N^{<z>}$ is provided with a
set of $\sigma $--matrices 
$$
\sigma _{\underleftarrow{i}\ \underrightarrow{i}}^{\underline{i}}=\sigma _{%
\underleftarrow{a_0}\ \underrightarrow{a_0}}^{\underline{a_0}}=l_i^{%
\underline{i}}\sigma _{\underleftarrow{i}\ \underrightarrow{i}}^i=l_{a_0}^{%
\underline{a_0}}\sigma _{\underleftarrow{a_0}\ \underrightarrow{a_0}}^{a_0}, 
$$
$$
\sigma _{\underleftarrow{a_1}\ \underrightarrow{a_1}}^{\underline{a_1}%
}=l_{a_1}^{\underline{a_1}}\sigma _{\underleftarrow{a_1}\ \underrightarrow{%
a_1}}^{a_1},...,\sigma _{\underleftarrow{a_p}\ \underrightarrow{a_p}}^{%
\underline{a_p}}=l_{a_p}^{\underline{a_p}}\sigma _{\underleftarrow{a_p}\ 
\underrightarrow{a_p}}^{a_p},...,\sigma _{\underleftarrow{a_z}\ 
\underrightarrow{a_z}}^{\underline{a_z}}=l_{a_z}^{\underline{a_z}}\sigma _{%
\underleftarrow{a_z}\ \underrightarrow{a_z}}^{a_z} 
$$
necessary for spinor parametizations of anticommuting variables%
$$
\theta ^{\underline{i}}=\zeta ^{\underline{a}_0}=(\theta ^{\underleftarrow{i}%
}=\zeta ^{\underleftarrow{a_0}},\theta ^{\underrightarrow{i}}=\zeta ^{%
\underrightarrow{a_0}}), 
$$
$$
\zeta ^{\underline{a_1}}=(\zeta ^{\underleftarrow{a_1}},\zeta ^{%
\underrightarrow{a_1}}),...,\zeta ^{\underline{a_p}}=(\zeta ^{%
\underleftarrow{a_p}},\zeta ^{\underrightarrow{a_p}}),...,\zeta ^{\underline{%
a_z}}=(\zeta ^{\underleftarrow{a_z}},\zeta ^{\underrightarrow{a_z}}) 
$$
(for simplicity, in sections 2--5 we consider Lorentz like spinor indices\\ $%
\left( ...,(\underleftarrow{a_p},\underrightarrow{a_p}),...\right) ,$ $%
p=0,1,2,...,z$ for a 4--dimensional even component of the s--space, but in
our case provided with N--connection structure).For simplicity, we shall
omit dots before and after indices enabled with subindex $p$ if this will
not give rise to ambiguities.

The locally adapted to N--connection partial s--derivations are introduced
in this manner:%
$$
\delta _{\underline{a_p}}=\frac \delta {\partial y^{\underline{a_p}%
}},~\delta _{\underleftarrow{a_p}}=\frac \delta {\partial \zeta ^{%
\underleftarrow{a_p}}}+i\sigma _{\underleftarrow{a_p}\ \underrightarrow{a_p}%
}^{a_p}\zeta ^{\underrightarrow{a_p}}\frac \delta {\partial y^{a_p}},~\delta
_{\underrightarrow{a_p}}=-\frac \delta {\partial \zeta ^{\underleftarrow{a_p}%
}}-i\zeta ^{\underleftarrow{a_p}}\sigma _{\underleftarrow{a_p}\ 
\underrightarrow{a_p}}^{a_p}{}\frac \delta {\partial y^{a_p}},\eqno(7) 
$$
where $i$ is the imaginary unity and, for instance, $\delta _{\underline{a_0}%
}=\frac \delta {\partial y^{\underline{a_0}}}=\delta _{\underline{i}}=\frac
\delta {\partial x^{\underline{i}}}$. La--derivations from (7) are
constructed by using operators (1). In brief we denote (7) as 
$$
\delta _{<\underline{\alpha }>}=l_{<\underline{\alpha }>}^{<\alpha >}\delta
_{<\alpha >}, 
$$
where matrix $l_{<\underline{\alpha }>}^{<\alpha >}$ is parametrized as 
$$
l_{<\underline{\alpha }>}^{<\alpha >}= 
$$
$$
\left( 
\begin{array}{cccccccc}
\delta _{\underline{i}}^i & 0 & 0 & ... & 0 & 0 & 0 & ... \\ 
i\sigma _{\underleftarrow{\widehat{i}}\ \underrightarrow{i}}^i\theta ^{%
\underrightarrow{i}} & \delta _{\underleftarrow{\widehat{i}}}^{%
\underleftarrow{i}} & 0 & ... & 0 & 0 & 0 & ... \\ 
-i\theta ^{\underleftarrow{i}}\sigma _{\underleftarrow{i}\ \underrightarrow{%
\widehat{j}}}^i\varepsilon ^{\underrightarrow{\widehat{j}}\underrightarrow{%
\widehat{i}}} & 0 & -\delta _{\underrightarrow{i}}^{\underrightarrow{%
\widehat{i}}} & ... & 0 & 0 & 0 & ... \\ 
... & ... & ... & ... & ... & ... & ... & ... \\ 
0 & 0 & 0 & ... & \delta _{\underline{a_p}}^{a_p} & 0 & 0 & ... \\ 
0 & 0 & 0 & ... & i\sigma _{\underleftarrow{\widehat{a}_p}\ \underrightarrow{%
a_p}}^{a_p}\zeta ^{\underrightarrow{a_p}} & \delta _{\underleftarrow{%
\widehat{a}_p}}^{\underleftarrow{a_p}} & 0 & ... \\ 
0 & 0 & 0 & ... & -i\zeta ^{\underleftarrow{a_p}}\sigma _{\underleftarrow{a_p%
}\ \underrightarrow{\widehat{b}_p}}^{a_p}\varepsilon ^{\underrightarrow{%
\widehat{b}_p}\underrightarrow{\widehat{a}_p}} & 0 & -\delta _{%
\underrightarrow{a_p}}^{\underrightarrow{\widehat{a}_p}} & ... \\ 
... & ... & ... & ... & ... & ... & ... & ...
\end{array}
\right) , 
$$
$\varepsilon $--objects of type $\varepsilon ^{\underrightarrow{\widehat{b}_p%
}\underrightarrow{\widehat{a}_p}}$ are spinor metrics. The inverse matrix $%
l_{<\alpha >}^{<\underline{\alpha }>}$ is introduced to satisfy conditions 
$$
l_{<\underline{\alpha }>}^{<\alpha >}l_{<\beta >}^{<\underline{\alpha }%
>}=\delta _{<\beta >}^{<\alpha >},~l_{<\alpha >}^{<\underline{\alpha }>}l_{<%
\underline{\beta }>}^{<\alpha >}=\delta _{<\underline{\beta }>}^{<\underline{%
\alpha }>}. 
$$
We call $l_{<\underline{\alpha }>}^{<\alpha >}~\left( l_{<\alpha >}^{<%
\underline{\alpha }>}\right) $ generalized supervielbein, s--vielbein,
(inverse s--viel\-be\-in) of N--connection s--space.

Additionally to (7) we shall use differential operators:%
$$
P_{a_p}=\frac \delta {\partial y^{\widehat{a}_p}},~Q_{\underleftarrow{a_p}%
}=\frac \delta {\partial \theta ^{\underleftarrow{a_p}}}-i\sigma _{%
\underleftarrow{a_p}\ \underrightarrow{a_p}}^{a_p}\zeta ^{\underrightarrow{%
a_p}}\frac \delta {\partial y^{a_p}},~Q_{\underrightarrow{a_p}}=-\frac
\delta {\partial \theta ^{\underleftarrow{a_p}}}+i\zeta ^{\underleftarrow{a_p%
}}\sigma _{\underleftarrow{a_p}\ \underrightarrow{a_p}}^{a_p}{}\frac \delta
{\partial y^{a_p}}.\eqno(8) 
$$
For a trivial N--connection operators (7) and (8) are transformed
respectively into ''covariant'' and infinitesimal generators on flat
s--spaces \cite{wes,mul,sal}.

\subsection*{2.3 Locally anisotropic superfields}

Functions on $u^{<\alpha >}$ are called superfields (s--fields) in $%
\widetilde{{\cal E}}_N^{<z>}.$ For simplicity we consider a real valued
s--field on dvs--bundle $\widetilde{{\cal E}}_N^{<1>}:$%
$$
V(u_{(p-1)},...,y_{(p)},\zeta _{(p)},...,y_{(z-1)},\zeta
_{(z-1)},y_{(z)},\zeta _{(z)})= 
$$
$$
V^{+}(u_{(p-1)},...,y_{(p)},\zeta _{(p)},...,y_{(z-1)},\zeta
_{(z-1)},y_{(z)},\zeta _{(z)}), 
$$
where by ''+'' is denoted the Hermitian conjugation. Every s--field is a
polynomial decomposition on variables $\left( \theta ^i,...,\zeta
^{a_p},...\right) .$ For instance, with respect to $(\zeta ^{\underleftarrow{%
a_z}},\zeta ^{\underrightarrow{a_z}})=(\underleftarrow{\zeta _{(z)}},%
\underrightarrow{\zeta _{(z)}})$ we have a such type polynom (for
simplicity, here we omit spinor indices):%
$$
V(u_{(z-1)},y_{(z)},\zeta ^{\underleftarrow{a_z}},\zeta ^{\underrightarrow{%
a_z}})=C(u_{(z-1)},y_{(z)})+i\underleftarrow{\zeta _{(z)}}\chi
(u_{(z-1)},y_{(z)})-i\underrightarrow{\zeta _{(z)}}\overline{\chi }%
(u_{(z-1)},y_{(z)})+ 
$$
$$
\frac i2\underleftarrow{\zeta _{(z)}}\underleftarrow{\zeta _{(z)}}(\mu
(u_{(z-1)},y_{(z)})+i\nu (u_{(z-1)},y_{(z)}))-\frac i2\underleftarrow{\zeta
_{(z)}}\underleftarrow{\zeta _{(z)}}(\mu (u_{(z-1)},y_{(z)})- 
$$
$$
i\nu (u_{(z-1)},y_{(z)}))-\underleftarrow{\zeta _{(z)}}\sigma ^i%
\underrightarrow{\zeta _{(z)}}v_i(u_{(z-1)},y_{(z)})+i\underleftarrow{\zeta
_{(z)}}\underleftarrow{\zeta _{(z)}}\underrightarrow{\zeta _{(z)}}\overline{%
\lambda }(u_{(z-1)},y_{(z)})- 
$$
$$
i\underrightarrow{\zeta _{(z)}}\underrightarrow{\zeta _{(z)}}\underleftarrow{%
\zeta _{(z)}}\lambda (u_{(z-1)},y_{(z)})-\frac 12\underleftarrow{\zeta _{(z)}%
}\underleftarrow{\zeta _{(z)}}\delta _i\chi (u_{(z-1)},y_{(z)})\sigma ^i%
\underrightarrow{\zeta _{(z)}}+ 
$$
$$
\frac 12\underrightarrow{\zeta _{(z)}}\underrightarrow{\zeta _{(z)}}%
\underleftarrow{\zeta _{(z)}}\sigma ^i\delta _i\overline{\chi }%
(u_{(z-1)},y_{(z)})+ 
$$
$$
\underleftarrow{\zeta _{(z)}}\underleftarrow{\zeta _{(z)}}\underleftarrow{%
\zeta _{(z)}}\underleftarrow{\zeta _{(z)}}\left( \frac
12P(u_{(z-1)},y_{(z)})+\frac 14(\delta ^I\delta
_I)C(u_{(z-1)},y_{(z)})\right) ,... 
$$
(in a similar manner we shall decompose on spinor variables\\ $(%
\underleftarrow{\zeta _{(z-1)}},\underrightarrow{\zeta _{(z-1)}}),...,(%
\underleftarrow{\zeta _{(1)}},\underrightarrow{\zeta _{(1)}}),(%
\underleftarrow{\theta },\underrightarrow{\theta })).$

\subsection*{2.4 Differential forms in N--connection s--spaces}

For locally adapted differentials (2) we introduce the supersymmetric
commutation rules%
$$
\delta u^{<\alpha >}\Lambda \delta u^{<\beta >}=-(-)^{|\alpha \beta |}\delta
u^{<\beta >}\Lambda \delta u^{<\alpha >}. 
$$
For every integer $q$ we introduce the linear space generated by basis
elements\\ $\delta u^{<\alpha _1>},\delta u^{<\alpha _2>},...,\delta
u^{<\alpha _q>}$ where every multiple satisfies the above presented
s--commutation rules. So, a function $\Phi (u)$ on $\widetilde{{\cal E}}%
_N^{<z>}$ is a 0--form, $\delta u^{<\alpha >}\Phi _{<\alpha >}$ is a
1--form, $\delta u^{<\alpha >}\Lambda \delta u^{<\beta >}\Phi _{<\alpha
><\beta >}$ is a 2--form and so on. Forms can be multiplied by taking into
account that $u^{<\alpha >}\delta u^{<\beta >}=(-)^{|\alpha \beta
|}u^{<\beta >}\delta u^{<\alpha >}.$

Let 
$$
\varpi =\delta u^{<\alpha _1>}\Lambda \delta u^{<\alpha _2>}\Lambda
...\Lambda \delta u^{<\alpha _q>} 
$$
be a q--form. The N--connection adapted differential%
$$
\delta \varpi =\delta u^{<\alpha _1>}\Lambda \delta u^{<\alpha _2>}\Lambda
...\Lambda \delta u^{<\alpha _q>}\Lambda \delta u^{<\alpha _{q+1}>}\frac{%
\delta \varpi }{\partial u^{<\alpha _{q+1}>}} 
$$
is a (q+1)--form. One holds the Poincare lemma: $\delta \delta =0.$ Under
some topological restrictions \cite{rog80,cia,leim} the inverse Poincare
lemma also holds: from $\delta \rho =0$ one follows that $\rho =\delta
\varpi $ (it should be noted here that on dvs--bundles we must take into
account the condition of existence of a N--connection structure).

An arbitrary locally adapted basis $l^{<\widehat{\alpha }>}$ in the space of
1--forms can be described by its s--vielbein matrix (generally being
different from the (8)):%
$$
\delta u^{<\alpha >}l_{<\alpha >}^{<\underline{\alpha }>}(u_{(z)})=l^{<%
\underline{\alpha }>}. 
$$
The inverse matrix $l_{<\underline{\alpha }>}^{<\alpha >}(u_{(z)})$ is
defined in a usual manner.

\section*{3. Higher Order Anisotropic Ga\-u\-ge S--Fields}

This section is devoted to the geometric background of gauge theory on
curved s--spaces provided with N--connection structure.

\subsection*{3.1 Gauge transforms in osculator s--bundles}

The structural group is a Lie group, acting on q--forms:%
$$
U=\exp \{i\Psi (u_{(z-1)},y_{(z)},\zeta ^{\underleftarrow{a_z}},\zeta ^{%
\underrightarrow{a_z}})\}, 
$$
$$
\Psi (u_{(z-1)},y_{(z)},\zeta ^{\underleftarrow{a_z}},\zeta ^{%
\underrightarrow{a_z}})=\sum\limits_{\widehat{e}}\gamma ^{\widehat{e}%
}(u_{(z-1)},y_{(z)},\zeta ^{\underleftarrow{a_z}},\zeta ^{\underrightarrow{%
a_z}})T_{\widehat{e}}, 
$$
where $T_{\widehat{e}}$ are generators of group :%
$$
\varsigma ^{\prime }=\varsigma U. 
$$
The connection form%
$$
\varphi =\delta u^{<\alpha >}\varphi _{<\alpha >}(u_{(z-1)},y_{(z)},\zeta ^{%
\underleftarrow{a_z}},\zeta ^{\underrightarrow{a_z}})=l^{<\underline{\alpha }%
>}\varphi _{<\underline{\alpha }>}(u_{(z-1)},y_{(z)},\zeta ^{\underleftarrow{%
a_z}},\zeta ^{\underrightarrow{a_z}})\eqno(9) 
$$
takes values in Lie algebra, i.e.%
$$
\varphi _{<\underline{\alpha }>}(u_{(z-1)},y_{(z)},\zeta ^{\underleftarrow{%
a_z}},\zeta ^{\underrightarrow{a_z}})=\sum\limits_{\widehat{e}}\varphi _{<%
\underline{\alpha }>}^{\widehat{e}}(u_{(z-1)},y_{(z)},\zeta ^{%
\underleftarrow{a_z}},\zeta ^{\underrightarrow{a_z}})T_{\widehat{e}}. 
$$
This s--field is a higher order anisotropic generalization of Yang--Mills
s--poten\-ti\-al. We have these transformation rules 
$$
\varphi ^{\prime }=U^{-1}\varphi U+U^{-1}\delta U, 
$$
$$
\varphi _{<\underline{\alpha }>}^{\prime }=U^{-1}\varphi _{<\underline{%
\alpha }>}U+U^{-1}\delta _{<\underline{\alpha }>}U, 
$$
where $\delta _{<\underline{\alpha }>}=l_{<\underline{\alpha }>}^{<\alpha
>}\delta _{<\alpha >}.$

The equation $K=\delta \varphi -\varphi \cdot \varphi $ shows that we can
construct ds--tensor values from connection s--potential (9). Computing
expressions 
$$
\delta \varphi =l^{<\underline{\alpha }>}\delta \varphi _{<\underline{\alpha 
}>}+\delta l^{<\underline{\alpha }>}~\varphi _{<\underline{\alpha }>}, 
$$
$$
\delta \varphi _{<\underline{\alpha }>}=l^{<\underline{\beta }>}\delta _{<%
\underline{\beta }>}\varphi _{<\underline{\alpha }>},\delta l^{<\underline{%
\alpha }>}=l^{<\underline{\beta }>}\delta _{<\underline{\beta }>}l^{<%
\underline{\alpha }>}\mbox{and}~\varphi \cdot \varphi =l^{<\underline{\alpha 
}>}l^{<\underline{\beta }>}\varphi _{<\underline{\alpha }>}\varphi _{<%
\underline{\beta }>}, 
$$
we get for coefficients of\\ $K=l^{<\underline{\alpha }>}\Lambda l^{<%
\underline{\beta }>}K_{<\underline{\alpha }><\underline{\beta }>},K_{<%
\underline{\beta }><\underline{\alpha }>}=-(-)^{|\underline{\alpha }%
\underline{\beta }|}K_{<\underline{\alpha }><\underline{\beta }>}$ the
formula:%
$$
K_{<\underline{\beta }><\underline{\alpha }>}=\delta _{<\underline{\beta }%
>}\varphi _{<\underline{\alpha }>}-(-)^{|\underline{\alpha }\underline{\beta 
}|}\delta _{<\underline{\alpha }>}\varphi _{<\underline{\beta }>}+ 
$$
$$
(-)^{|<\underline{\beta }>(<\underline{\alpha }>+<\underline{\mu }>)|}l_{<%
\underline{\alpha }>}^{<\alpha >}(\delta _{<\underline{\beta }>}l_{<\alpha
>}^{<\underline{\gamma }>})\varphi _{<\underline{\gamma }>}-(-)^{|<%
\underline{\beta }><\underline{\mu }>|}l_{<\underline{\beta }>}^{<\alpha
>}(\delta _{<\underline{\alpha }>}l_{<\alpha >}^{<\underline{\gamma }%
>})\varphi _{<\underline{\gamma }>}- 
$$
$$
\varphi _{<\underline{\beta }>}\varphi _{<\underline{\alpha }>}+(-)^{|%
\underline{\alpha }\underline{\beta }|}\varphi _{<\underline{\alpha }%
>}\varphi _{<\underline{\beta }>}+w_{<\underline{\beta }><\underline{\alpha }%
>}^{<\gamma >}\varphi _{<\gamma >}, 
$$
where $w_{<\underline{\beta }><\underline{\alpha }>}^{<\gamma >}$ unholonomy
coefficients are defined as in (3). For example, we present the structure of
ds--components of the coefficients :%
$$
K_{a_pb_p}=\delta _{a_p}\varphi _{b_p}-\delta _{b_p}\varphi _{a_p}+[\varphi
_{a_p},\varphi _{b_p}]+w_{a_pb_p}^{<\alpha >}\varphi _{<\alpha >},~ 
$$
$$
K_{a_p\underleftarrow{b_p}}=\delta _{a_p}\varphi _{\underleftarrow{b_p}%
}-\delta _{\underleftarrow{b_p}}\varphi _{a_p}+[\varphi _{a_p},\varphi _{%
\underleftarrow{b}}]+w_{a_p\underleftarrow{b_p}}^{<\alpha >}\varphi
_{<\alpha >}, 
$$
$$
K_{a_p\underrightarrow{b_p}}=\delta _{a_p}\varphi _{\underrightarrow{b_p}%
}-\delta _{\underrightarrow{b_p}}\varphi _{a_p}+[\varphi _{a_p},\varphi _{%
\underrightarrow{b_p}}]+w_{a_p\underrightarrow{b_p}}^{<\alpha >}\varphi
_{<\alpha >},~ 
$$
$$
K_{\underleftarrow{a_p}\underleftarrow{b_p}}=\delta _{\underleftarrow{a_p}%
}\varphi _{\underleftarrow{b_p}}+\delta _{\underleftarrow{b_p}}\varphi _{%
\underleftarrow{a_p}}+\{\varphi _{\underleftarrow{a_p}},\varphi _{%
\underleftarrow{b_p}}\}+w_{\underleftarrow{a_p\ }\underleftarrow{b_p}%
}^{<\alpha >}\varphi _{<\alpha >}, 
$$
$$
K_{\underrightarrow{a_p}\underrightarrow{b_p}}=\delta _{\underrightarrow{a_p}%
}\varphi _{\underrightarrow{b_p}}+\delta _{\underrightarrow{b_p}}\varphi _{%
\underrightarrow{a_p}}+\{\varphi _{\underrightarrow{a_p}},\varphi _{%
\underrightarrow{b_p}}\}+w_{\underrightarrow{a_p}\underrightarrow{b_p}%
}^{<\alpha >}\varphi _{<\alpha >}, 
$$
$$
K_{\underleftarrow{a_p}\ \underrightarrow{b_p}}=\delta _{\underleftarrow{a_p}%
}\varphi _{\underrightarrow{b_p}}+\delta _{\underrightarrow{b_p}}\varphi _{%
\underleftarrow{a_p}}+2i\sigma _{\underleftarrow{a_p}\ \underrightarrow{b_p}%
}^{\widehat{a}_p}\varphi _{\widehat{a}_p}+\{\varphi _{\underleftarrow{a_p}%
},\varphi _{\underrightarrow{b_p}}\}+w_{\underleftarrow{a_p}\ 
\underrightarrow{b_p}}^{<\alpha >}\varphi _{<\alpha >}, 
$$
where $p=0,1,2,...,z.$

In the next subsections we shall analyze constraints substantially
decreasing the number of independent fields containing components of
s--field $\varphi _{<\widehat{\alpha }>}$ without restrictions on theirs
dependence on coordinates $u^{<\alpha >}.$

\subsection*{3.2 Abelian locally anisotropic s--fields}

This class of s--fields satisfy conditions $[\varphi _{<\alpha >}\varphi
_{<\beta >}\}=0.$ If constraints 
$$
K_{\underleftarrow{a_p}\underleftarrow{b_p}}=\delta _{\underleftarrow{a_p}%
}\varphi _{\underleftarrow{b_p}}+\delta _{\underleftarrow{b_p}}\varphi _{%
\underleftarrow{a_p}}+w_{\underleftarrow{a_p\ }\underleftarrow{b_p}%
}^{<\alpha >}\varphi _{<\alpha >},K_{\underrightarrow{a_p}\underrightarrow{%
b_p}}=\delta _{\underrightarrow{a_p}}\varphi _{\underrightarrow{b_p}}+\delta
_{\underrightarrow{b_p}}\varphi _{\underrightarrow{a_p}}+w_{\underrightarrow{%
a_p}\underrightarrow{b_p}}^{<\alpha >}\varphi _{<\alpha >} 
$$
are imposed, there are such s--fields $A\left( u\right) ,B\left( u\right) $
that 
$$
\varphi _{\underleftarrow{a_p}}=-i\delta _{\underleftarrow{a_p}}A,~\varphi _{%
\underrightarrow{a_p}}=\delta _{\underrightarrow{a_p}}B. 
$$
With respect to gauge shifts by a s--function $\kappa (u)$%
$$
A\rightarrow A+\kappa +S^{+},~S\mbox{ is a s--function satisfying}\ \delta _{%
\underrightarrow{i}}S=0,...,\delta _{\underrightarrow{a_p}}S=0, 
$$
$$
B\rightarrow B-\kappa +T,\ T\mbox{ is a s--function satisfying}\ \delta _{%
\underrightarrow{i}}T=0,...,\delta _{\underrightarrow{a_p}}T=0 
$$
one holds these transformation laws:%
$$
\varphi _{\underleftarrow{a_p}}\rightarrow \varphi _{\underleftarrow{a_p}%
}-i\delta _{\underleftarrow{a_p}}\kappa ,\varphi _{\underrightarrow{a_p}%
}=\varphi _{\underrightarrow{a_p}}-i\delta _{\underrightarrow{a_p}}\kappa . 
$$
If (additionally) equations%
$$
K_{\underleftarrow{a_p}\ \underrightarrow{b_p}}=\delta _{\underleftarrow{a_p}%
}\varphi _{\underrightarrow{b_p}}+\delta _{\underrightarrow{b_p}}\varphi _{%
\underleftarrow{a_p}}+2i\sigma _{\underleftarrow{a_p}\ \underrightarrow{b_p}%
}^{\widehat{a}_p}\varphi _{\widehat{a}_p}+w_{\underleftarrow{a_p}\ 
\underrightarrow{b_p}}^{<\alpha >}\varphi _{<\alpha >}=0\eqno(10) 
$$
are satisfied, the s--functions $\varphi _{\widehat{i}},...,\varphi _{%
\widehat{a}_p},...$ can be expressed through $A$ and $B;$ gauge transforms
of these fields are parametrized as 
$$
\varphi _{<\alpha >}\rightarrow \varphi _{<\alpha >}-i\delta _{<\alpha
>}\kappa . 
$$

So we can express s--field $\varphi _{<\alpha >},$ as well curvatures\\ $%
K_{ij},...,K_{a_pb_p},...,K_{i\underleftarrow{j}},K_{i\underrightarrow{j}%
},...,K_{a_p\underleftarrow{b_p}},K_{a_p\underrightarrow{b_p}},...$ as
functions on $A,B.$ Really, all invariants can be expressed through values
these values: 
$$
W_{\underleftarrow{a_p}}=\varepsilon ^{\underrightarrow{b_p}\ 
\underrightarrow{c_p}}\delta _{\underrightarrow{b_p}}\delta _{%
\underrightarrow{c_p}}\delta _{\underleftarrow{a_p}}(A+B),...,W_{%
\underrightarrow{a_p}}=\varepsilon ^{\underleftarrow{b_p}\underleftarrow{c_p}%
}\delta _{\underleftarrow{b_p}}\delta _{\underleftarrow{c_p}}\delta _{%
\underrightarrow{a_p}}(A+B), 
$$
$$
K_{a_p\underleftarrow{b_p}}=\frac 18\sigma _{a_p}^{\underrightarrow{c_p}\ 
\underleftarrow{d_p}}\varepsilon _{\underleftarrow{d_p}\ \underleftarrow{b_p}%
}W_{\underrightarrow{c_p}},K_{a_p\underrightarrow{b_p}}=-\frac 18\varepsilon
_{\underrightarrow{b_p}\ \underrightarrow{c_p}}\sigma _{a_p}^{%
\underrightarrow{c_p}\ \underleftarrow{d_p}}W_{\underleftarrow{d_p}}, 
$$
$$
K_{a_pb_p}=\frac i{64}\{(\varepsilon \sigma _{a_p}\sigma _{b_p})^{%
\underleftarrow{c_p}\ \underleftarrow{d_p}}(\delta _{\underleftarrow{d_p}}W_{%
\underleftarrow{c_p}}+ 
$$
$$
\delta _{\underleftarrow{c_p}}W_{\underleftarrow{d_p}})+(\varepsilon \sigma
_{a_p}\sigma _{b_p})^{\underrightarrow{c_p}\ \underrightarrow{d_p}}(\delta _{%
\underrightarrow{c_p}}W_{\underrightarrow{d_p}}+\delta _{\underrightarrow{d_p%
}}W_{\underrightarrow{c_p}})\}\ . 
$$
From definition of $W_{\underleftarrow{i}},W_{\underrightarrow{i}},...,W_{%
\underleftarrow{a_p}},W_{\underrightarrow{a_p}},...$ one follows that%
$$
\delta _{\underrightarrow{d_p}}W_{\underleftarrow{c_p}}=0,\delta _{%
\underleftarrow{c_p}}W_{\underrightarrow{d_p}}=0,\delta ^{\underleftarrow{a_p%
}}W_{\underleftarrow{a_p}}-\delta _{\underrightarrow{b_p}}W^{%
\underrightarrow{b_p}}=0. 
$$

The reality conditions for our theory are specified by conditions $\varphi
=\varphi ^{+}$ within a gauge transform when $A+B=(A+B)^{+}$ $W_{%
\underrightarrow{i}}$ and $S=T.$ In this case the corresponding Lagrangian
is chosen as 
$$
L\sim W^{\underleftarrow{i}}W_{\underleftarrow{i}}+W^{\underleftarrow{a_1}%
}W_{\underleftarrow{a_1}}+...+W^{\underleftarrow{a_z}}W_{\underleftarrow{a_z}%
}+W^{\underrightarrow{i}}W_{\underrightarrow{i}}+W^{\underrightarrow{a_1}}W_{%
\underrightarrow{a_p}}+...+W^{\underrightarrow{a_z}}W_{\underrightarrow{a_z}%
}.\eqno(11) 
$$

In a similar manner we consider nonabelian gauge s--fields of s--spaces with
local anisotropy.

\subsection*{3.3 Nonabelian locally anisotropic gauge s--fields}

Constraints are imposed as in the Abelian case:%
$$
K_{\underleftarrow{a_p}\ \underleftarrow{b_p}}=K_{\underrightarrow{a_p}\ 
\underrightarrow{b_p}}=K_{\underleftarrow{a_p}\ \underrightarrow{b_p}}=0. 
$$
From $K_{\underleftarrow{i}\ \underleftarrow{j}}=...=K_{\underleftarrow{a_p}%
\ \underleftarrow{b_p}}=...0$ and $K_{\underrightarrow{i}\ \underrightarrow{j%
}}=...=K_{\underrightarrow{a_p}\ \underrightarrow{b_p}}=...0$ we
respectively obtain 
$$
\varphi _{\underleftarrow{a_p}}=-e^{-A}\delta _{\underleftarrow{a_p}}e^A%
\eqno(12) 
$$
and%
$$
\varphi _{\underrightarrow{a_p}}=-e^{-A}\delta _{\underrightarrow{a_p}}e^A. 
$$
Considering transforms 
$$
e^A\rightarrow e^{S^{+}}e^Ae^\kappa ~\mbox{and}\ e^{-U}\rightarrow
e^{-T}e^{-A}e^\kappa ,\eqno(13) 
$$
from which one follows transformation laws for $A$ and $B:$%
$$
A\rightarrow A+\kappa +S^{+}+...\ \mbox{and}\ B\rightarrow B-\kappa +T+..., 
$$
and imposing constraints of type (10) we can express, similarly as in
Abelian case, the s--field $\varphi _{<\alpha >},$ and curvatures\\ $%
K_{ij},...,K_{a_pb_p},...,K_{i\underleftarrow{j}},K_{i\underrightarrow{j}%
},...,K_{a_p\underleftarrow{b_p}},K_{a_p\underrightarrow{b_p}},...$ as
functions on $A,B,$ and, as a consequence, as functions of invariants $W_{%
\underleftarrow{i}},W_{\underrightarrow{i}},...,W_{\underleftarrow{a_p}},W_{%
\underrightarrow{a_p}}.$

Finally, in this subsection we remark that traces\\ $Tr(W^{\underleftarrow{i}%
}W_{\underleftarrow{i}}),Tr(W^{\underleftarrow{a_1}}W_{\underleftarrow{a_1}%
}),...,Tr(W^{\underleftarrow{a_z}}W_{\underleftarrow{a_z}})$ are gauge
invariant and%
$$
\delta _{\underrightarrow{i}}Tr(W^{\underleftarrow{i}}W_{\underleftarrow{i}%
})=...=\delta _{\underrightarrow{a_p}}Tr(W^{\underleftarrow{a_p}}W_{%
\underleftarrow{a_p}})=...=0. 
$$
So, we can use these traces and theirs complex conjugations in order to
define Lagrangians of type (11) for nonabelian gauge theories. Reality
conditions and gauge transforms can be considered as in the Abelian case but
by taking into account changing (12) and (13).

\section*{4. Su\-per\-gra\-vi\-ty in Osculator S--Bundles}

The generalized s--vielbein $E_{<\alpha >}^{<\underline{\alpha }>}$ and
connection form $\Phi _{<\underline{\beta }><\mu >}^{<\underline{\alpha }>},$
the last takes values in a Lie algebra, are considered as basic variables on
osculator bundle $Osc^z\widetilde{M}$ with base $\widetilde{M}$ being of
dimension $\left( (3,1),1\right) $ where $\left( 3,1\right) $ denotes
respectively the dimension and signature of the even subspace and $1$ is the
dimention of the odd subspace. Our aim is to find respectively a higher
order extension of d--covariant equations for the field of spin 2 and spin
3/2. As a possible structural group we choose, for instance, the Lorentz
subgroup (locally the action of this group is split according to the fixed
N--connection structure). With respect to coordinate d--transforms 
$$
\delta u^{<\mu ^{\prime }>}=\delta u^{<\nu >}\frac{\delta u^{<\mu ^{\prime
}>}}{\partial u^{<\nu >}} 
$$
one holds the transformation laws%
$$
E_{<\alpha ^{\prime }>}^{<\underline{\alpha }>}=E_{<\alpha >}^{<\underline{%
\alpha }>}\frac{\delta u^{<\alpha >}}{\partial u^{<\alpha ^{\prime }>}}\ 
\mbox{and}\ \Phi _{<\underline{\beta }><\mu ^{\prime }>}^{<\underline{\alpha 
}>}=\frac{\delta u^{<\nu >}}{\partial u^{<\mu ^{\prime }>}}\Phi _{<%
\underline{\beta }><\nu >}^{<\underline{\alpha }>}. 
$$

Let introduce 1--forms 
$$
E^{<\underline{\alpha }>}=E_{<\mu >}^{<\underline{\alpha }>}\delta u^{<\mu
>}\ \mbox{and}\ \Phi _{<\underline{\beta }>}^{<\underline{\alpha }>}=\Phi _{<%
\underline{\beta }><\mu >}^{<\underline{\alpha }>}\delta u^{<\mu >} 
$$
satisfying transformation laws of type:%
$$
E^{<\underline{\alpha ^{\prime }}>}=E^{<\underline{\alpha }>}X_{<\underline{%
\alpha }>}^{<\underline{\alpha }^{\prime }>}\ \mbox{and}\ \Phi _{<\underline{%
\beta }^{\prime }>}^{<\underline{\alpha }^{\prime }>}=X_{\quad <\underline{%
\beta }^{\prime }>}^{-1\ <\underline{\alpha }>}\Phi _{<\underline{\alpha }%
>}^{<\underline{\beta }>}X_{<\underline{\beta }>}^{<\underline{\alpha }%
^{\prime }>}+X_{\quad <\underline{\beta }^{\prime }>}^{-1\ <\underline{%
\alpha }>}\delta X_{<\underline{\alpha }>}^{<\underline{\alpha }^{\prime
}>}. 
$$

The torsion and curvature are defined respectively by the first and second
structure equations%
$$
\Omega ^{<\underline{\alpha }>}=\delta E^{<\underline{\alpha }>}-E^{<%
\underline{\beta }>}\Phi _{<\underline{\beta }>}^{<\underline{\alpha }>}, 
$$
and 
$$
R_{<\underline{\alpha }>}^{<\underline{\beta }>}=\delta \Phi _{<\underline{%
\alpha }>}^{<\underline{\beta }>}-\Phi _{<\underline{\alpha }>}^{<\underline{%
\gamma }>}\Phi _{<\underline{\gamma }>}^{<\underline{\beta }>}. 
$$
The coefficients with values in Lie algebra are written in this form:%
$$
\Omega _{<\underline{\beta }><\underline{\gamma }>}^{<\underline{\alpha }%
>}=(-)^{|<\underline{\beta }>(<\underline{\gamma }>+<\underline{\mu }>)|}E_{<%
\underline{\gamma }>}^{<\mu >}E_{<\underline{\beta }>}^{<\nu >}\delta _{<\nu
>}E_{<\mu >}^{<\underline{\alpha }>}- 
$$
$$
(-)^{|<\underline{\gamma }><\underline{\mu }>)|}E_{<\underline{\beta }%
>}^{<\mu >}E_{<\underline{\gamma }>}^{<\nu >}\delta _{<\nu >}E_{<\mu >}^{<%
\underline{\alpha }>}-\Phi _{<\underline{\gamma }><\underline{\beta }>}^{<%
\underline{\alpha }>}+(-)^{|<\underline{\beta }><\underline{\gamma }>|}\Phi
_{<\underline{\beta }><\underline{\gamma }>}^{<\underline{\alpha }>}\eqno(14)
$$
and 
$$
R_{<\underline{\delta }><\underline{\varepsilon }><\underline{\alpha }>}^{<%
\underline{\beta }>}=(-)^{|<\underline{\delta }>(<\underline{\varepsilon }>+<%
\underline{\mu }>)|}E_{<\underline{\varepsilon }>}^{<\mu >}E_{<\underline{%
\delta }>}^{<\nu >}\delta _{<\nu >}\Phi _{<\underline{\alpha }><\mu >}^{<%
\underline{\beta }>}- 
$$
$$
(-)^{|<\underline{\varepsilon }><\underline{\mu }>|}E_{<\underline{\delta }%
>}^{<\mu >}E_{<\underline{\varepsilon }>}^{<\nu >}\delta _{<\nu >}\Phi _{<%
\underline{\alpha }><\mu >}^{<\underline{\beta }>}- 
$$
$$
(-)^{|<\underline{\delta }>(<\underline{\varepsilon }>+<\underline{\alpha }%
>+<\underline{\gamma }>)|}\Phi _{<\underline{\alpha }><\underline{%
\varepsilon }>}^{<\underline{\gamma }>}\Phi _{<\underline{\gamma }><%
\underline{\delta }>}^{<\underline{\beta }>}+(-)^{|<\underline{\varepsilon }%
>(<\underline{\alpha }>+<\underline{\gamma }>)|}\Phi _{<\underline{\alpha }><%
\underline{\delta }>}^{<\underline{\gamma }>}\Phi _{<\underline{\gamma }><%
\underline{\varepsilon }>}^{<\underline{\beta }>}. 
$$

Putting $E^{<\underline{\alpha }>}=l^{<\underline{\alpha }>},$ with $%
l_{<\alpha >}^{<\underline{\alpha }>}$ from (8), $\Phi _{<\underline{\beta }%
><\underline{\gamma }>}^{<\underline{\alpha }>}=0$ in (14) and for a
vanishing N--connection we obtain that torsion for a trivial osculator
s--space has components:%
$$
\Omega _{\underleftarrow{b_p}\ \underrightarrow{c_p}}^{(0)\underline{a_p}%
}=\Omega _{\underrightarrow{b_p}\ \underleftarrow{c_p}}^{(0)\underline{a_p}%
}=2i\sigma _{\underleftarrow{b_p}\ \underrightarrow{c_p}}^{\underline{a_p}%
}=0,\eqno(15) 
$$
the rest of components are zero.

In order to consider the linearized osculator s--gravity we substitute $%
E_{<\alpha >}^{<\underline{\alpha }>}=l_{<\alpha >}^{<\underline{\alpha }%
>}+kh_{<\alpha >}^{<\underline{\alpha }>}$ in (8), where $k$ is the
interaction constant and $h_{<\alpha >}^{<\underline{\alpha }>}$ is linear
perturbation of a d--frame in N--connection s--space. By straightforward
calculations we can verify that from equations (15) and 
$$
\Omega _{\underline{b_p}\underline{c_p}}^{\underline{a_p}}=\Omega _{%
\underleftarrow{b_p}\ \underleftarrow{c_p}}^{\underleftarrow{a_p}}=\Omega _{%
\underleftarrow{b_p}\ \underleftarrow{c_p}}^{\underrightarrow{a_p}}=\Omega _{%
\underrightarrow{b_p}\ \underrightarrow{c_p}}^{\underleftarrow{a_p}}=\Omega
_{\underleftarrow{b_p}\ \underleftarrow{c_p}}^{\underline{a_p}}=\Omega _{%
\underleftarrow{b_p}\ \underline{c_p}}^{\underline{a_p}}=0 
$$
one follows only algebraic relations. In this case on the base s--space of
the osculator s--bundle we obtain exactly a dynamical system of equations
for spin 2 and spin 3/2 fields (for usual supergravity see \cite
{des,free,gris,brei,west,sal} ). We can also solve nonlinear equations. It
is convenient to introduce the special gauge when for $\underleftarrow{%
\theta }=\underrightarrow{\theta }=...=\underleftarrow{\zeta _{(p)}}=%
\underrightarrow{\zeta _{(p)}}=...=0$ the s--vielbein and d--connection are
prametrized 
$$
E_i^{\underline{i}}=l_i^{\underline{i}}\left( u\right) ,...,E_{a_p}^{%
\underline{a_p}}=l_{a_p}^{\underline{a_p}}\left( u_{(p-1)},x_{(p)}\right)
,...,E_{\underleftarrow{i}}^{\underleftarrow{\widehat{i}}}=\delta _{%
\underleftarrow{i}}^{\underleftarrow{\widehat{i}}},...,E_{\underleftarrow{a_p%
}}^{\underleftarrow{\widehat{a}_p}}=\delta _{\underleftarrow{a_p}}^{%
\underleftarrow{\widehat{a}_p}},...,...\eqno(16) 
$$
$$
E_i^{\underleftarrow{i}}=\frac 12\psi _i^{\underleftarrow{i}%
}(x),...,E_{a_p}^{\underleftarrow{a_p}}=\frac 12\psi _{a_p}^{\underleftarrow{%
a_p}}(u_{(p-1)},y_{(p)}),..., 
$$
$$
E_{i\underrightarrow{i}}=\frac 12\psi _{i\underrightarrow{j}%
}(u_{(p-1)},x_{(p)}),...,E_{a_p\underrightarrow{a_p}}=\frac 12\psi _{a_p%
\underrightarrow{a_p}}(u_{(p-1)},y_{(p)}),..., 
$$
the rest of components of the s--vielbein are zero, and 
$$
\Phi _{\underleftarrow{j}\ k}^{\underleftarrow{i}}=\varphi _{\underleftarrow{%
j}\ k}^{\underleftarrow{i}}(x),...,\Phi _{\underleftarrow{b_p}\ c_p}^{%
\underleftarrow{a_p}}=\varphi _{\underleftarrow{b_p}\ c_p}^{\underleftarrow{%
a_p}}(u_{(p-1)},y_{(p)}),...,\eqno(17) 
$$
the rest of components of d--connection are zero.

Fields $l_i^{\underline{i}}\left( u\right) ,...,l_{a_p}^{\underline{a_p}%
}\left( u_{(p-1)}\right) ,...,\psi _i^{\underleftarrow{i}}(x),...,\psi
_{a_p}^{\underleftarrow{a_p}}(u_{(p-1)},y_{(p)}),...$ and\\ $\varphi _{%
\underleftarrow{j}\ k}^{\underleftarrow{i}}(x),...\varphi _{\underleftarrow{%
b_p}\ c_p}^{\underleftarrow{a_p}}(u_{(p-1)},y_{(p)}),...$ from (16) and (17)
are corresponding extensions of the tetrad, Rarita--Schwinger and connection
fields on osculator bundle \cite{mirata}.

We note that equations 
$$
\Omega _{\underleftarrow{i}\ \underline{k}}^{\underrightarrow{j}}=\Omega _{%
\underleftarrow{i}\ \underline{k}}^{\underrightarrow{j}}=0\ \left(
...,\Omega _{\underleftarrow{b_p}\ \underline{c_p}}^{\underrightarrow{a_p}%
}=\Omega _{\underleftarrow{b_p}\ \underline{c_p}}^{\underrightarrow{a_p}%
}=0,...\right) 
$$
defines the dynamics in $x$--space ( in $\left( u_{(p-1)},x_{(p)}\right) $%
--space).The rest of nonvanishing components of torsion are computed by
putting components (16) and (17) into (14) :%
$$
\Omega _{ij\mid \underleftarrow{\theta }=\underrightarrow{\theta }=0}^{%
\widehat{k}}={\cal T}_{ij}^{\widehat{k}}=\frac i2(\psi _i\sigma ^{\widehat{k}%
}\overline{\psi }_j-\psi _j\sigma ^{\widehat{k}}\overline{\psi }_i),..., 
$$
$$
\Omega _{b_pc_p\mid \underleftarrow{\zeta _{(p)}}=\underrightarrow{\zeta
_{(p)}}=0}^{\widehat{a}_p}={\cal T}_{b_pc_p}^{\widehat{a}_p}=\frac i2(\psi
_{b_p}\sigma ^{\widehat{k}}\overline{\psi }_{c_p}-\psi _{c_p}\sigma ^{%
\widehat{k}}\overline{\psi }_{b_p}),..., 
$$
which shows that torsion in dvs--bundles can be generated by a corresponding
distribution of spin density, and 
$$
\Omega _{ij\mid \underleftarrow{\theta }=\underrightarrow{\theta }=0}^{%
\underleftarrow{i}}={\cal T}_{ij}^{\underleftarrow{i}}=\frac 12(D_i\psi _j^{%
\underleftarrow{i}}-D_j\psi _i^{\underleftarrow{i}}),..., 
$$
$$
\Omega _{b_pc_p\mid \underleftarrow{\zeta _{(p)}}=\underrightarrow{\zeta
_{(p)}}=0}^{\underleftarrow{a_p}}={\cal T}_{b_pc_p}^{\underleftarrow{a_p}%
}=\frac 12(D_{b_p}\psi _{c_p}^{\underleftarrow{a_p}}-D_{c_p}\psi _{b_p}^{%
\underleftarrow{i}}),..., 
$$
where $D_i,...,D_{b_p},...$ are usual d--covariant derivatives on the base
of osculator space.

We note that in this and next sections we shall omit tedious calculations
being similar to those from \cite{west}; we shall present the final results
and emphasize that we can verify them in a straightforward manner by taking
into account the distinguished character of geometical objects and the
interactions with the N--connection fields.

The Bianchi identities in the osculator s--bundle are written as 
$$
\delta \Omega ^{<\underline{\alpha }>}+\Omega ^{<\underline{\beta }>}\Phi _{<%
\underline{\beta }>}^{<\underline{\alpha }>}-E^{<\underline{\beta }>}R_{~<%
\underline{\beta }>}^{<\underline{\alpha }>}=0, 
$$
$$
\delta R_{~<\underline{\alpha }>}^{<\underline{\beta }>}+R_{~<\underline{%
\alpha }>}^{<\underline{\gamma }>}R_{~<\underline{\gamma }>}^{<\underline{%
\beta }>}-\Phi _{<\underline{\alpha }>}^{<\underline{\gamma }>}\Phi _{<%
\underline{\gamma }>}^{<\underline{\beta }>}=0, 
$$
or, in coefficient form, as 
$$
E^{<\underline{\gamma }>}E^{<\underline{\beta }>}E^{<\underline{\alpha }%
>}(E_{<\underline{\alpha }>}^{<\mu >}{\cal D}_{<\mu >}\Omega _{<\underline{%
\beta }><\underline{\gamma }>}^{<\underline{\delta }>}+ 
$$
$$
\Omega _{<\underline{\alpha }><\underline{\beta }>}^{<\underline{\tau }%
>}\Omega _{<\underline{\tau }><\underline{\gamma >}}^{<\underline{\delta }%
>}-R_{<\underline{\alpha }><\underline{\beta }><\underline{\gamma }>}^{<%
\underline{\delta }>}=0,\eqno(18) 
$$
$$
E^{<\underline{\gamma }>}E^{<\underline{\beta }>}E^{<\underline{\alpha }%
>}\left( E_{<\underline{\alpha }>}^{<\mu >}{\cal D}_{<\mu >}R_{<\underline{%
\beta }><\underline{\gamma }><\underline{\delta }>}^{<\underline{\tau }%
>}+\Omega _{<\underline{\alpha }><\underline{\beta }>}^{<\underline{%
\varepsilon }>}R_{<\underline{\varepsilon }><\underline{\gamma }><\underline{%
\delta }>}^{<\underline{\tau }>}\right) =0, 
$$
where the supersymmetric gauge d--covariant derivation ${\cal D}_{<\mu >}$
acts, for instance, as 
$$
{\cal D}_{<\mu >}X^{<\underline{\alpha }>}=\delta _{<\mu >}X^{<\underline{%
\alpha }>}+(-)^{|<\underline{\beta }><\mu >|}X^{<\underline{\beta }>}\Phi _{<%
\underline{\beta }><\mu >}^{<\underline{\alpha }>}, 
$$
and%
$$
{\cal D}_{<\mu >}X_{<\underline{\alpha }>}=\delta _{<\mu >}X_{<\underline{%
\alpha }>}-\Phi _{<\underline{\alpha }><\mu >}^{<\underline{\beta }>}X_{<%
\underline{\beta }>}. 
$$

Introducing parametrizations (16) (the special gauge) in (18) we find
equations:%
$$
\sigma _{\underleftarrow{b_p}\ \underrightarrow{c_p}}^{\underline{a_p}}l_{%
\underline{a_p}}^{d_p}l_{\underline{e_p}}^{e_p}\left( D_{d_p}\overline{\psi }%
_{e_p}^{\underrightarrow{c_p}}-D_{e_p}\overline{\psi }_{d_p}^{%
\underrightarrow{c_p}}\right) =0\eqno(19) 
$$
(the Rarita--Shwinger equations on osculator bundle) and%
$$
{\cal R}_{\underline{a_p}\ \underline{c_p}\ \underline{b_p}}^{\underline{b_p}%
}+il^{\underline{b_p}\ d_p}\psi _{d_p}^{\underleftarrow{e_p}}\sigma _{%
\underline{a_p}\ \underleftarrow{e_p}\ \underrightarrow{f_p}}{\cal T}_{%
\underline{b_p}\ \underline{c_p}}^{\underrightarrow{f_p}}+il^{\underline{b_p}%
d_p}\overline{\psi }_{d_p}^{\underrightarrow{f_p}}\sigma _{\underline{a_p}\ 
\underleftarrow{e_p}\ \underrightarrow{f_p}}{\cal T}_{\underline{b_p}\ 
\underline{c_p}}^{\underleftarrow{e_p}}=0,\eqno(20) 
$$
where components of sourse are 
$$
{\cal T}_{\underline{a_p}\ \underline{b_p}}^{\underrightarrow{c_p}}=\frac
12l_{\underline{a_p}}^{a_p}l_{\underline{b_p}}^{b_p}(D_{a_p}\psi _{b_p}^{%
\underrightarrow{c_p}}-D_{b_p}\psi _{a_p}^{\underrightarrow{c_p}}), 
$$
and curvature (in the special gauge)\ is expressed as%
$$
{\cal R}_{\underline{a_p}\underline{b_p}c_pd_p}=l_{c_p}^{\underline{e_p}%
}l_{d_p}^{\underline{f_p}}R_{\underline{a_p}\underline{b_p}\underline{e_p}%
\underline{f_p}}-i(l_{c_p}^{\underline{f_p}}\psi _{d_p}^{\underleftarrow{s_p}%
}-l_{d_p}^{\underline{f_p}}\psi _{c_p}^{\underleftarrow{s_p}})\sigma _{%
\underline{f_p}\ \underleftarrow{s_p}\ \underrightarrow{t_p}}\Omega _{%
\underline{a_p}\ \underline{b_p}}^{\underrightarrow{t_p}}- 
$$
$$
i(l_{d_p}^{\underline{f_p}}\overline{\psi }_{c_p\underrightarrow{s_p}%
}-l_{c_p}^{\underline{f_p}}\overline{\psi }_{d_p\underrightarrow{s_p}%
})\sigma _{\underline{f_p}}^{\underleftarrow{t_p}\ \underrightarrow{s_p}%
}\Omega _{\underline{a_p}\ \underline{b_p}\ \underleftarrow{t_p}\mid 
\underleftarrow{\zeta _{(p)}}=\underrightarrow{\zeta _{(p)}}=0}, 
$$
$$
R_{\underline{b_p}c_pd_p}^{\underline{a_p}}=E_{c_p}^{\underline{c_p}%
}E_{d_p}^{\underline{d_p}}R_{\underline{b_p}\underline{c_p}\underline{d_p}}^{%
\underline{a_p}}+E_{c_p}^{\underleftarrow{s_p}}E_{d_p}^{\underline{d_p}}R_{%
\underline{b_p}\underleftarrow{s_p}\underline{d_p}}^{\underline{a_p}}+ 
$$
$$
E_{c_p}^{\underline{c_p}}E_{d_p}^{\underleftarrow{s_p}}R_{\underline{b_p}%
\underleftarrow{s_p}\underline{c_p}}^{\underline{a_p}}+E_{c_p%
\underrightarrow{t_p}}E_{d_p}^{\underline{d_p}}R_{.\underline{b_p}.%
\underline{d_p}}^{\underline{a_p}.\underrightarrow{t_p}}+E_{c_p}^{\underline{%
c_p}}E_{d_p\underrightarrow{t_p}}R_{\underline{b_p}\underline{c_p}}^{%
\underline{a_p}..\underrightarrow{t_p}}= 
$$
$$
\delta _{d_p}\Phi _{\underline{b_p}c_p}^{\underline{a_p}}-\delta _{c_p}\Phi
_{\underline{b_p}d_p}^{\underline{a_p}}+\Phi _{\underline{b_p}d_p}^{%
\underline{f_p}}\Phi _{\underline{f_p}c_p}^{\underline{a_p}}-\Phi _{%
\underline{b_p}c_p}^{\underline{f_p}}\Phi _{\underline{f_p}d_p}^{\underline{%
a_p}}+w_{c_pd_p}^{e_p}\Phi _{e_p\underline{b_p}}^{\underline{a_p}}. 
$$

Finally, in this section, we note that for trivial N--connection structures 
on vector superbundles the equations (19) and (20) are transformed into
dynamical field equations for the model of supergravity developed by S.\
Deser and B. Zumino \cite{des}.

\section*{5. Bianchi Identities in Osculat\-or S--Bundles}

The purpose of this section is the analyzes of Bianchi identities in the
framework of locally anisotropic supergravity theory on osculator s--bundle.
We shall impose s--gravitational constraints and solve these identities with
respect to s--fields and theirs ds--covariant derivations.

\subsection*{5.1 Distinguished Bianchi identities}

The Bianchi identities are written as (18). Constraints on torsion are
imposed in general form as (16) with the rest of components being zero.By
using technique developed in $\cite{grimm}$ we shall solve in explicit form
the system (18) with the mentioned type of constraints on osculator
s--bundle in explicit form. We note that to do this we shall not use an
explicit form of ds--covariant derivation; the necessary information is
contained in the s--symmetric commutator%
$$
[{\cal D}_{<\alpha >},{\cal D}_{<\beta >}\}=-R_{~\bullet <\alpha ><\beta
>}^{\bullet }-\Omega _{<\alpha ><\beta >}^{<\gamma >}{\cal D}_{<\gamma >}. 
$$

In order to find solutions we distinguish identities (18) in this form:%
$$
R_{\underleftarrow{a_p}\underleftarrow{b_p}\underleftarrow{c_d}%
\underleftarrow{ep}}+R_{\underleftarrow{b_p}\underleftarrow{c_d}%
\underleftarrow{a_p}\underleftarrow{ep}}+R_{\underleftarrow{c_d}%
\underleftarrow{a_p}\underleftarrow{b_p}\underleftarrow{ep}}=0,\eqno(21) 
$$

$$
R_{\underleftarrow{a_p}\underrightarrow{b_p}\underleftarrow{c_p}%
\underleftarrow{d_p}}+R_{\underrightarrow{b_p}\underleftarrow{c_p}%
\underleftarrow{a_p}\underleftarrow{d_p}}+2i\sigma _{\underleftarrow{c_p}%
\underrightarrow{b_p}}^{e_p}\Omega _{\underleftarrow{a_p}e_p\underleftarrow{%
d_p}}+2i\sigma _{\underleftarrow{a_p}\underrightarrow{b_p}}^{e_p}\Omega _{%
\underleftarrow{c_p}e_p\underleftarrow{d_p}}=0,\eqno(22) 
$$

$$
R_{\underleftarrow{a_p}\underleftarrow{b_p}\underrightarrow{c_p}%
\underrightarrow{d_p}}=-2i\sigma _{\underleftarrow{b_p}\underrightarrow{c_p}%
}^{e_p}\Omega _{\underleftarrow{a_p}e_p\underrightarrow{d_p}}-2i\sigma _{%
\underleftarrow{a_p}\underrightarrow{c_p}}^{e_p}\Omega _{\underleftarrow{b_p}%
e_p\underrightarrow{d_p}},\eqno(23) 
$$

$$
R_{\underrightarrow{a_p}\underrightarrow{b_p}\underline{c_p}\underline{d_p}%
}=-2i\sigma _{\underline{d_p}\underleftarrow{s_p}\underrightarrow{b_p}%
}\Omega _{\underrightarrow{a_p}\underline{c_p}}^{\underleftarrow{s_p}%
}-2i\sigma _{\underline{d_p}\underleftarrow{s_p}\underrightarrow{a_p}}\Omega
_{\underrightarrow{b_p}\underline{c_p}}^{\underleftarrow{s_p}},\eqno(24) 
$$

$$
R_{\underleftarrow{a_p}\underrightarrow{b_p}\underline{c_p}\underline{d_p}%
}=-2i\sigma _{\underline{d_p}\underleftarrow{s_p}\underrightarrow{b_p}%
}\Omega _{\underleftarrow{a_p}\underline{c_p}}^{\underleftarrow{s_p}%
}-2i\sigma _{\underline{d_p}\underleftarrow{s_p}\underleftarrow{a_p}}\Omega
_{\underrightarrow{b_p}\underline{c_p}}^{\underleftarrow{s_p}},\eqno(25) 
$$

$$
R_{\underline{a_p}\underline{b_p}\underline{c_p}\underline{d_p}}+R_{%
\underline{b_p}\underline{c_p}\underline{a_p}\underline{d_p}}+R_{\underline{%
c_p}\underline{a_p}\underline{b_p}\underline{d_p}}=0,\eqno(26) 
$$
(linear equations without derivatives)%
$$
R_{\underleftarrow{a_p}\underline{b_p}\underleftarrow{c_p}\underleftarrow{d_p%
}}+R_{\underleftarrow{c}\underline{b_p}\underleftarrow{a_p}\underleftarrow{%
d_p}}+{\cal D}_{\underleftarrow{c_p}}\Omega _{\underleftarrow{a_p}\underline{%
b_p}\underleftarrow{d_p}}+{\cal D}_{\underleftarrow{a_p}}\Omega _{%
\underleftarrow{c_p}\underline{b_p}\underleftarrow{d_p}}=0,\eqno(27) 
$$

$$
R_{\underline{a_p}\underleftarrow{b_p}\underrightarrow{c_p}\underrightarrow{%
d_p}}={\cal D}_{\underleftarrow{b_p}}\Omega _{\underrightarrow{c_p}%
\underline{a_p}\underrightarrow{d_p}}+{\cal D}_{\underrightarrow{c_p}}\Omega
_{\underleftarrow{b_p}\underline{a_p}\underrightarrow{d_p}}+2i\sigma _{%
\underleftarrow{b_p}\underrightarrow{c_p}}^{m_p}\Omega _{m_p\underline{a_p}%
\underrightarrow{d_p}},\eqno(28) 
$$

$$
{\cal D}_{\underrightarrow{k}}\Omega _{\underrightarrow{j}\underline{i}%
\underrightarrow{l}}+{\cal D}_{\underrightarrow{j}}\Omega _{\underrightarrow{%
k}\underline{i}\underrightarrow{l}}=0,\eqno(29) 
$$
(linear identities containing derivations) and%
$$
R_{\underline{a_p}\underline{b_p}\underrightarrow{c_p}\underrightarrow{d_p}}=%
{\cal D}_{\underline{a_p}}\Omega _{\underline{b_p}\underrightarrow{c_p}%
\underrightarrow{d_p}}+{\cal D}_{\underline{b_p}}\Omega _{\underrightarrow{%
c_p}\underline{a_p}\underrightarrow{d_p}}+{\cal D}_{\underrightarrow{c_p}%
}\Omega _{\underline{a_p}\underline{b_p}\underrightarrow{d_p}}+\eqno(30) 
$$
$$
\Omega _{\underline{b_p}\underrightarrow{c_p}}^{\underleftarrow{m_p}}\Omega
_{\underleftarrow{m_p}\underline{a_p}\underrightarrow{d_p}}+\Omega _{%
\underline{b_p}\underrightarrow{c_p}}^{\underrightarrow{m_p}}\Omega _{%
\underrightarrow{m_p}\underline{a_p}\underrightarrow{d_p}}+\Omega _{%
\underrightarrow{c_p}\underline{a_p}}^{\underleftarrow{m_p}}\Omega _{%
\underleftarrow{m_p}\underline{b_p}\underrightarrow{d_p}}+\Omega _{%
\underrightarrow{c_p}\underline{a_p}}^{\underrightarrow{m_p}}\Omega _{%
\underrightarrow{m_p}\underline{b_p}\underrightarrow{d_p}},..., 
$$

$$
R_{\underline{a_p}\underline{b_p}\underrightarrow{c_p}\underrightarrow{d_p}}=%
{\cal D}_{\underline{a_p}}\Omega _{\underline{b_p}\underrightarrow{c_p}%
\underleftarrow{d_p}}+{\cal D}_{\underline{b_p}}\Omega _{\underrightarrow{c_p%
}\underline{a_p}\underleftarrow{d_p}}+{\cal D}_{\underrightarrow{c_p}}\Omega
_{\underline{a_p}\underline{b_p}\underleftarrow{d_p}}+\eqno(31) 
$$
$$
\Omega _{\underline{b_p}\underrightarrow{c_p}}^{\underleftarrow{m_p}}\Omega
_{\underleftarrow{m_p}\underline{a_p}\underleftarrow{d_p}}+\Omega _{%
\underline{b_p}\underrightarrow{c_p}}^{\underrightarrow{m_p}}\Omega _{%
\underrightarrow{m_p}\underline{a_p}\underleftarrow{d_p}}+\Omega _{%
\underrightarrow{c_p}\underline{a_p}}^{\underleftarrow{m_p}}\Omega _{%
\underleftarrow{m_p}\underline{b_p}\underleftarrow{d_p}}+\Omega _{%
\underrightarrow{c_p}\underline{a_p}}^{\underrightarrow{m_p}}\Omega _{%
\underrightarrow{m_p}\underline{b_p}\underleftarrow{d_p}}=0,..., 
$$

$$
{\cal D}_{\underline{a_p}}\Omega _{\underline{b_p}\underline{c_p}%
\underleftarrow{d_p}}+{\cal D}_{\underline{b_p}}\Omega _{\underline{c_p}%
\underline{a_p}\underleftarrow{d_p}}+{\cal D}_{\underline{c_p}}\Omega _{%
\underline{a_p}\underline{b_p}\underleftarrow{d_p}}+\Omega _{\underline{a_p}%
\underline{b_p}}^{\underleftarrow{m_p}}\Omega _{\underleftarrow{m_p}%
\underline{c_p}\underleftarrow{d_p}}+\Omega _{\underline{a_p}\underline{b_p}%
}^{\underrightarrow{m_p}}\Omega _{\underrightarrow{m_p}\underline{c_p}%
\underleftarrow{d_p}}+\eqno(32) 
$$
$$
\Omega _{\underline{b_p}\underline{c_p}}^{\underleftarrow{m_p}}\Omega _{%
\underleftarrow{m_p}\underline{a_p}\underleftarrow{d_p}}+\Omega _{\underline{%
b_p}\underline{c_p}}^{\underrightarrow{m_p}}\Omega _{\underrightarrow{m_p}%
\underline{a_p}\underleftarrow{d_p}}+\Omega _{\underline{c_p}\underline{a_p}%
}^{\underleftarrow{m_p}}\Omega _{\underleftarrow{m_p}\underline{b_p}%
\underleftarrow{d_p}}+\Omega _{\underline{c_p}\underline{a_p}}^{%
\underrightarrow{m_p}}\Omega _{\underrightarrow{m_p}\underline{b_p}%
\underleftarrow{d_p}},...~ 
$$
(nonlinear identities).

For a trivial osculator s--bundle $Osc^0\widetilde{M}=\widetilde{M},\dim 
\widetilde{M}=(4,1),\widetilde{M}$ being a s--symmetric extension of the
Lorentz bundle formulas (21)--(32) are transformed into the Bianchi
identities for the locally isotropic s--gravity model considered by \cite
{grimm}.

\subsection*{5.2 Solution of distinguished Bianchi identities}

It is convenient to use spinor decompositions of curvatures and torsions
(see, for instance, \cite{pen,penr1,penr2}) 
$$
R_{\underrightarrow{a_p}\underrightarrow{b_p}\underleftarrow{c_p}%
\underrightarrow{d_p}\underleftarrow{e_p}\underrightarrow{f_p}}=\sigma _{%
\underleftarrow{c_p}\underrightarrow{d_p}}^{d_p}\sigma _{\underleftarrow{e_p}%
\underrightarrow{f_p}}^{f_p}R_{\underrightarrow{a_p}\underrightarrow{b_p}%
d_pf_p},\quad \Omega _{\underrightarrow{a_p}\underleftarrow{b_p}%
\underrightarrow{c_p}\underleftarrow{d_p}}=\sigma _{\underleftarrow{b_p}%
\underrightarrow{c_p}}^{e_p}\Omega _{\underrightarrow{a_p}e_p\underleftarrow{%
d_p}} 
$$
and 
$$
R_{\underrightarrow{a_p}\underrightarrow{b_p}\underleftarrow{c_p}%
\underrightarrow{d_p}\underleftarrow{e_p}\underrightarrow{f_p}%
}=-2\varepsilon _{\underleftarrow{c_p}\underleftarrow{e_p}}R_{%
\underrightarrow{a_p}\underrightarrow{b_p}\underrightarrow{d_p}%
\underrightarrow{f_p}}+2\varepsilon _{\underrightarrow{d_p}\underrightarrow{%
f_p}}R_{\underrightarrow{a_p}\underrightarrow{b_p}\underleftarrow{c_p}%
\underleftarrow{e_p}},...\ . 
$$

Let start with the solutions of linear equations without derivatives. By
straightforward calculations we can verify that 
$$
\Omega _{\underrightarrow{i}\underleftarrow{j}\underrightarrow{k}%
\underleftarrow{l}}=-2i\varepsilon _{\underrightarrow{i}\underrightarrow{k}%
}\varepsilon _{\underleftarrow{j}\underleftarrow{l}}R_{(0)},...,\Omega _{%
\underrightarrow{a_p}\underleftarrow{b_p}\underrightarrow{c_p}%
\underleftarrow{d_p}}=-2i\varepsilon _{\underrightarrow{a_p}\underrightarrow{%
c_p}}\varepsilon _{\underleftarrow{b_p}\underleftarrow{d_p}}R_{(p)},...%
\eqno(33) 
$$
and 
$$
R_{\underrightarrow{i}\underrightarrow{j}\underrightarrow{k}\underrightarrow{%
l}}=4\left( \varepsilon _{\underrightarrow{i}\underrightarrow{l}}\varepsilon
_{\underrightarrow{j}\underrightarrow{k}}+\varepsilon _{\underrightarrow{j}%
\underrightarrow{l}}\varepsilon _{\underrightarrow{i}\underrightarrow{k}%
}\right) R_{(0)},..., 
$$
$$
R_{\underrightarrow{a_p}\underrightarrow{b_p}\underrightarrow{c_p}%
\underrightarrow{d_p}}=4\left( \varepsilon _{\underrightarrow{a_p}%
\underrightarrow{d_p}}\varepsilon _{\underrightarrow{b_p}\underrightarrow{c_p%
}}+\varepsilon _{\underrightarrow{b_p}\underrightarrow{d_p}}\varepsilon _{%
\underrightarrow{a_p}\underrightarrow{c_p}}\right) R_{(p)},..., 
$$
where 
$$
R_{(0)}=g^{kl}R_{kli}^i,...,R_{(p)}=g^{a_pc_p}R_{a_pc_pb_p}^{b_p},..., 
$$
satisfy correspondingly identities (2.18) and (2.19).

Similarly we can check that spinor decompositions%
$$
\Omega _{\underleftarrow{a_p}\underleftarrow{b_p}\underrightarrow{c_p}%
\underleftarrow{d_p}}=\frac i4\left( \varepsilon _{\underleftarrow{b_p}%
\underleftarrow{d_p}}Q_{\underleftarrow{a_p}\underrightarrow{c_p}%
}-3\varepsilon _{\underleftarrow{a_p}\underleftarrow{b_p}}Q_{\underleftarrow{%
d_p}\underrightarrow{c_p}}-3\varepsilon _{\underleftarrow{a_p}%
\underleftarrow{d_p}}Q_{\underleftarrow{b_p}\underrightarrow{c_p}}\right) , 
$$
$$
R_{\underleftarrow{a_p}\underrightarrow{b_p}\underleftarrow{c_p}%
\underleftarrow{d_p}=}\varepsilon _{c_pa_p}Q_{\underleftarrow{d_p}%
\underrightarrow{b_p}}+\varepsilon _{d_pa_p}Q_{\underleftarrow{c_p}%
\underrightarrow{b_p}}, 
$$
$$
Q_{\underleftarrow{a_p}\underrightarrow{b_p}}^{*}=Q_{\underleftarrow{a_p}%
\underrightarrow{b_p}} 
$$
solve identities (21),(22) and (23).

The identity (26) allows us to express a part of curvature components by
using  some components of torsion: 
$$
R_{\underleftarrow{a_p}\underline{b_p}\underline{c_p}\underline{d_p}%
}=i(\sigma _{\underline{b_p}\underleftarrow{a_p}\underrightarrow{e_p}}\Omega
_{\underline{c_p}\underline{d_p}}^{\underrightarrow{e_p}}-\sigma _{%
\underline{d_p}\underleftarrow{a_p}\underrightarrow{e_p}}\Omega _{\underline{%
b_p}\underline{c_p}}^{\underrightarrow{e_p}}-\sigma _{\underline{c_p}%
\underleftarrow{a_p}\underrightarrow{e_p}}\Omega _{\underline{d_p}\underline{%
b_p}}^{\underrightarrow{e_p}}). 
$$

Now we consider these spinor decompositions of curvatures:%
$$
R_{\underleftarrow{a_p}\underrightarrow{a_p}\underleftarrow{b_p}%
\underrightarrow{b_p}\underleftarrow{c_p}\underrightarrow{c_p}%
\underleftarrow{d_p}\underrightarrow{d_p}}=\varepsilon _{\underleftarrow{a_p}%
\underleftarrow{b_p}}\varepsilon _{\underleftarrow{c_p}\underleftarrow{d_p}}%
\overline{\chi }_{\underrightarrow{a_p}\underrightarrow{b_p}\underrightarrow{%
c_p}\underrightarrow{d_p}}+\varepsilon _{\underrightarrow{a_p}%
\underrightarrow{b_p}}\varepsilon _{\underrightarrow{c_p}\underrightarrow{d_p%
}}\chi _{\underleftarrow{a_p}\underleftarrow{b_p}\underleftarrow{c_p}%
\underleftarrow{d_p}}- 
$$
$$
\varepsilon _{\underleftarrow{a_p}\underleftarrow{b_p}}\varepsilon _{%
\underrightarrow{c_p}\underrightarrow{d_p}}\overline{\varphi }_{%
\underrightarrow{a_p}\underrightarrow{b_p}\underleftarrow{c_p}%
\underleftarrow{d_p}}-\varepsilon _{\underrightarrow{a_p}\underrightarrow{b_p%
}}\varepsilon _{\underleftarrow{c_p}\underleftarrow{d_p}}\varphi _{%
\underleftarrow{a_p}\underleftarrow{b_p}\underrightarrow{c_p}%
\underrightarrow{d_p}}. 
$$
The identity (21) is satisfied if 
$$
\varphi _{\underleftarrow{a_p}\underleftarrow{b_p}\underrightarrow{c_p}%
\underrightarrow{d_p}}=\overline{\varphi }_{\underrightarrow{c_p}%
\underrightarrow{d_p}\underleftarrow{a_p}\underleftarrow{b_p}},\chi _{\quad 
\underleftarrow{b_p}\underleftarrow{a_p}\underleftarrow{c_p}}^{%
\underleftarrow{a_p}}=\varepsilon _{\underleftarrow{b_p}\underleftarrow{c_p}%
}\Lambda ~(\Lambda \text{~\mbox{ is real}}). 
$$

The next step is the solution of linear identities (28),(29) and (30)
containing derivatives. Putting (32) into (28) we find 
$$
{\cal D}_{\underrightarrow{a_p}}R_{(p)}=0. 
$$
The d--spinor%
$$
\Omega _{\underrightarrow{e_p}\underleftarrow{a_p}\underrightarrow{b_p}%
\underleftarrow{c_p}\underrightarrow{d_p}}=\sigma _{\underleftarrow{a_p}%
\underrightarrow{b_p}}^{\underline{a_p}}\sigma _{\underleftarrow{c_p}%
\underrightarrow{d_p}}^{\underline{b_p}}\Omega _{\underrightarrow{e_p}%
\underline{a_p}\underline{b_p}} 
$$
can be decomposed into irreducible parts \cite{pen,penr1}%
$$
\Omega _{\underrightarrow{e_p}\underleftarrow{a_p}\underrightarrow{b_p}%
\underleftarrow{c_p}\underrightarrow{d_p}}=-\varepsilon _{\underleftarrow{a_p%
}\underleftarrow{c_p}}(W_{\underrightarrow{b_p}\underrightarrow{c_p}%
\underleftarrow{e_p}}+\varepsilon _{\underrightarrow{e_p}\underrightarrow{d_p%
}}\tau _{\underrightarrow{b_p}}+\varepsilon _{\underrightarrow{e_p}%
\underrightarrow{b_p}}\tau _{\underrightarrow{d_p}})+\varepsilon _{%
\underrightarrow{a_p}\underrightarrow{c_p}}\tau _{\underleftarrow{a_p}%
\underleftarrow{c_p}\underrightarrow{e_p}}, 
$$
where $W_{\underrightarrow{b_p}\underrightarrow{c_p}\underleftarrow{e_p}}$
is an arbitrary d--spinor and $\tau _{\underrightarrow{b_p}}$ and $\tau _{%
\underleftarrow{a_p}\underleftarrow{c_p}\underrightarrow{e_p}}$ are
expressed through derivations of $Q_{\underleftarrow{a_p}\underrightarrow{b_p%
}}$ (see below ). A tedious but trivial calculus can convince us that the
solution of (28) can be expressed as 
$$
\Omega _{\underrightarrow{e_p}\underleftarrow{a_p}\underrightarrow{b_p}%
\underleftarrow{c_p}\underrightarrow{d_p}}=-\varepsilon _{\underleftarrow{a_p%
}\underleftarrow{c_p}}W_{\underrightarrow{b_p}\underrightarrow{c_p}%
\underleftarrow{e_p}}+\frac 12\varepsilon _{\underrightarrow{a_p}%
\underrightarrow{c_p}}({\cal D}_{\underleftarrow{a_p}}Q_{\underleftarrow{c_p}%
\underrightarrow{e_p}}+{\cal D}_{\underleftarrow{c_p}}Q_{\underleftarrow{b_p}%
\underrightarrow{e_p}})- 
$$
$$
\frac 12\varepsilon _{\underleftarrow{a_p}\underleftarrow{c_p}}(\varepsilon
_{\underrightarrow{e_p}\underrightarrow{d_p}}{\cal D}^{\underleftarrow{f_p}%
}Q_{\underleftarrow{f_p}\underrightarrow{b_p}}+\varepsilon _{%
\underrightarrow{e_p}\underrightarrow{b_p}}{\cal D}^{\underleftarrow{f_p}}Q_{%
\underleftarrow{f_p}\underrightarrow{d_p}}), 
$$
$$
R_{\underleftarrow{a_p}\underleftarrow{b_p}\underrightarrow{c_p}%
\underleftarrow{d_p}\underleftarrow{e_p}}=\frac i2(\varepsilon _{%
\underleftarrow{a_p}\underleftarrow{b_p}}{\cal D}_{\underleftarrow{d_p}}Q_{%
\underleftarrow{e_p}\underrightarrow{c_p}}+\varepsilon _{\underleftarrow{a_p}%
\underleftarrow{d_p}}{\cal D}_{\underleftarrow{b_p}}Q_{\underleftarrow{e_p}%
\underrightarrow{c_p}}+ 
$$
$$
\varepsilon _{\underleftarrow{a_p}\underleftarrow{b_p}}{\cal D}_{%
\underleftarrow{e_p}}Q_{\underleftarrow{d_p}\underrightarrow{c_p}%
}+\varepsilon _{\underleftarrow{a_p}\underleftarrow{d_p}}{\cal D}_{%
\underleftarrow{e_p}}Q_{\underleftarrow{b_p}\underrightarrow{c_p}%
})+i(\varepsilon _{\underleftarrow{d_p}\underleftarrow{a_p}}\varepsilon _{%
\underleftarrow{b_p}\underleftarrow{e_p}}+\varepsilon _{\underleftarrow{e_p}%
\underleftarrow{a_p}}\varepsilon _{\underleftarrow{b_p}\underleftarrow{d_p}})%
{\cal D}^{\underleftarrow{f_p}}Q_{\underleftarrow{f_p}\underrightarrow{c_p}%
}, 
$$
$$
R_{\underleftarrow{a_p}\underleftarrow{b_p}\underrightarrow{c_p}%
\underrightarrow{d_p}\underrightarrow{e_p}}=2i\varepsilon _{\underleftarrow{%
a_p}\underleftarrow{b_p}}W_{\underrightarrow{d_p}\underrightarrow{e_p}%
\underrightarrow{c_p}}+\frac i2(\varepsilon _{\underrightarrow{c_p}%
\underrightarrow{d_p}}{\cal D}_{\underleftarrow{a_p}}Q_{\underleftarrow{b_p}%
\underrightarrow{e_p}}+\varepsilon _{\underrightarrow{c_p}\underrightarrow{%
e_p}}{\cal D}_{\underleftarrow{a_p}}Q_{\underleftarrow{b_p}\underrightarrow{%
d_p}}). 
$$
This solution is compatible with (29) if 
$$
{\cal D}^{c_p}Q_{\underleftarrow{c_p}\underrightarrow{e_p}}={\cal D}_{%
\underleftarrow{e_p}}R_{(p)}^{*}. 
$$

So we have solved all linear identities.

Nonlinear relations (30),(31) and (32) can be transformed into linear ones
by using commutators of d--covariant derivations. Omitting such algebraic
transforms we present expressions 
$$
\overline{\chi }_{\underrightarrow{a_p}\underrightarrow{b_p}\underrightarrow{%
c_p}\underrightarrow{d_p}}=\frac 14({\cal D}_{\underrightarrow{a_p}}W_{%
\underrightarrow{b_p}\underrightarrow{c_p}\underrightarrow{d_p}}+{\cal D}_{%
\underrightarrow{b_p}}W_{\underrightarrow{c_p}\underrightarrow{d_p}%
\underrightarrow{a_p}}+{\cal D}_{\underrightarrow{c_p}}W_{\underrightarrow{%
d_p}\underrightarrow{a_p}\underrightarrow{b_p}}+{\cal D}_{\underrightarrow{%
d_p}}W_{\underrightarrow{a_p}\underrightarrow{b_p}\underrightarrow{c_p}})+ 
$$
$$
(\varepsilon _{\underrightarrow{a_p}\underrightarrow{d_p}}\varepsilon _{%
\underrightarrow{c_p}\underrightarrow{b_p}}+\varepsilon _{\underrightarrow{%
b_p}\underrightarrow{d_p}}\varepsilon _{\underrightarrow{c_p}%
\underrightarrow{a_p}})[\frac 1{16}(\overline{{\cal D}}_{\underrightarrow{e_p%
}}\overline{{\cal D}}^{\underrightarrow{e_p}}R_{(p)}^{*}+{\cal D}^{%
\underleftarrow{e_p}}{\cal D}_{\underleftarrow{e_p}}R_{(p)})+\frac 18Q_{%
\underleftarrow{e_p}\underrightarrow{e_p}}Q^{\underleftarrow{e_p}%
\underrightarrow{e_p}}-2RR^{*}] 
$$
and 
$$
\varphi _{\underleftarrow{a_p}\underleftarrow{b_p}\underrightarrow{c_p}%
\underrightarrow{d_p}}=\overline{\varphi }_{\underrightarrow{c_p}%
\underrightarrow{d_p}\underleftarrow{a_p}\underleftarrow{b_p}}=\frac 14(Q_{%
\underleftarrow{a_p}\underrightarrow{d_p}}Q_{\underleftarrow{b_p}%
\underrightarrow{c_p}}+Q_{\underleftarrow{b_p}\underrightarrow{d_p}}Q_{%
\underleftarrow{a_p}\underrightarrow{c_p}})+ 
$$
$$
\frac i8({\cal D}_{\underleftarrow{b_p}\underrightarrow{c_p}}Q_{%
\underleftarrow{a_p}\underrightarrow{d_p}}+{\cal D}_{\underleftarrow{a_p}%
\underrightarrow{c_p}}Q_{\underleftarrow{b_p}\underrightarrow{d_p}}+{\cal D}%
_{\underleftarrow{b_p}\underrightarrow{d_p}}Q_{\underleftarrow{a_p}%
\underrightarrow{c_p}}+{\cal D}_{\underleftarrow{a_p}\underrightarrow{d_p}%
}Q_{\underleftarrow{b_p}\underrightarrow{c_p}}+ 
$$
$$
{\cal D}_{\underrightarrow{c_p}}{\cal D}_{\underleftarrow{b_p}}Q_{%
\underleftarrow{a_p}\underrightarrow{d_p}}+{\cal D}_{\underrightarrow{c_p}}%
{\cal D}_{\underleftarrow{a_p}}Q_{\underleftarrow{b_p}\underrightarrow{d_p}}+%
{\cal D}_{\underrightarrow{d_p}}{\cal D}_{\underleftarrow{b_p}}Q_{%
\underleftarrow{a_p}\underrightarrow{c_p}}+{\cal D}_{\underrightarrow{d_p}}%
{\cal D}_{\underleftarrow{a_p}}Q_{\underleftarrow{b_p}\underrightarrow{c_p}%
}) 
$$
which solve (30); if conditions 
$$
{\cal D}_{\underrightarrow{a_p}}W_{\underrightarrow{b_p}\underrightarrow{c_p}%
\underrightarrow{d_p}}=0 
$$
are satisfied we obtain solutions (31) and (32).

\section*{6. Einstein--Cartan Structures in DVS--Bun\-dles}

In this section we shall introduce a set of Einstein like gravitational
equations, i.e. we shall formulate a variant of higher order anisotropic
supergravity on dsv-bundle ${\cal E}^{<z>}$ over a supersmooth manifold $%
\widetilde{M}$. This model will contain as particular cases the Miron and
Anastasiei locally anisotropic gravity \cite{ana86,ma87,ma94} on vector
bundles (they considered prescribed components of N-connection and h(hh)-
and v(vv)--torsions; in our supesymmetric approach \cite{vlasg} we used
algebraic equations for torsion and its source in order to close the system
of field equations). There are two ways in developing supergravitational
models. We can try to maintain similarity to Einstein's general relativity,
see in \cite{arn,nat} an example of such type locally isotropic
supergravity, and to formulate a variant of Einstein--Cartan theory on
dvs--bundles (this will be the purpose of this section) or to introduce into
consideration generalized supervielbein variables and to formulate a
supersymmetric gauge like model of la-supergravity (this approach is more
accepted in the usual locally isotropic supergravity, see as reviews \cite
{sal,west,mul}). The second variant will be analyzed in the next section by
using the s-bundle of supersymmetric affine adapted frames on la-superspaces.

Let consider a dvs--bundle ${\cal E}^{<z>}$ provided with some compatible
nonlinear connection N, d--connection $D$ and metric $G$ structures (see 
conventions and details in subsection 1.3 and \cite{vjph2}). For a locally
N-adapted frame (1) we write  the components of a d-connection $D$ (4) as 
$$
{{{\Gamma }^I}_{JK}}={{L^I}_{JK}},{{{\Gamma }^I}_{J<A>}}={{C^I}_{J<A>}},{{%
\Gamma }^I}_{<A>J}=0,{{\Gamma }^I}_{<A><B>}=0,\eqno(34) 
$$
$$
{{\Gamma }^{<A>}}_{JK}=0,{{\Gamma }^{<A>}}_{J<B>}=0,{{\Gamma }^{<A>}}_{<B>K}=%
{L^{<A>}}_{<B>K}, 
$$
$$
{{{\Gamma }^{<A>}}_{<B><C>}}={{C^{<A>}}_{<B><C>}}. 
$$
The nonholonomy coefficients ${{w^{<\gamma >}}_{<\alpha ><\beta >}}$ (see
(3)) of d--connection (34) are computed as follows: 
$$
{{w^K}_{IJ}}=0,{{w^K}_{<A>J}}=0,{{w^K}_{I<A>}}=0,{{w^K}_{<A><B>}}=0,{{w^{<A>}%
}_{IJ}}={R^{<A>}}_{IJ}, 
$$
$$
{{w^{<B>}}_{<A>I}}=-(-)^{|I<A>|}{\frac{\partial {N_I^B}}{\partial y^{<A>}}},{%
{w^{<B>}}_{I<A>}}={\frac{\partial {N_I^{<B>}}}{\partial y^{<A>}}},{{w^{<C>}}%
_{<A><B>}}=0. 
$$
By straightforward calculations we obtain respectively these components of
torsion, ${\cal T}({\delta }_{<\gamma >},{\delta }_{<\beta >})={{\cal T}%
_{\cdot <\beta ><\gamma >}^{<\alpha >}}{\delta }_{<\alpha >},$ and
curvature, ${\cal R}({\delta }_{<\beta >},{\delta }_{<\gamma >}){\delta }%
_{<\tau >}={{\cal R}_{.<\beta ><\gamma ><\tau >}^{<\alpha >}}{{\delta }%
_{<\alpha >}},$ ds-tensors: 
$$
{\cal T}_{\cdot JK}^I={T^I}_{JK},{\cal T}_{\cdot J<A>}^I={C^I}_{J<A>},{\cal T%
}_{\cdot J<A>}^I=-{C^I}_{J<A>},{\cal T}_{\cdot <A><B>}^I=0, 
$$
$$
{\cal T}_{\cdot IJ}^{<A>}={R^{<A>}}_{IJ},{\cal T}_{\cdot I<B>}^{<A>}=-{%
P^{<A>}}_{<B>I},{\cal T}_{\cdot <B>I}^{<A>}={P^{<A>}}_{<B>I}, 
$$
$$
{\cal T}_{\cdot <B><C>}^{<A>}={S^{<A>}}_{<B><C>} 
$$
and 
$$
{\cal R}_{\cdot IKL}^J=R_{IKL}^J,{\cal R}_{\cdot BKL}^J=0,{\cal R}_{\cdot
JKL}^{<A>}=0,{\cal R}_{\cdot <B>KL}^{<A>}={R}_{\cdot <B>KL}^{<A>}, 
$$
$$
{\cal R}_{J\cdot K<D>}^I={{P_J}^I}_{K<D>},{\cal R}_{<B>K<D>}^I=0,{\cal R}%
_{JK<D>}^{<A>}=0, 
$$
$$
{\cal R}_{<B>K<D>}^{<A>}={P}_{<B>K<D>}^{<A>},{\cal R}_{J<D>K}^I=-(-)^{|<D>K|}%
{P}_{JK<D>}^I, 
$$
$$
{\cal R}_{<B><D>K}^I=0,{\cal R}_{J<D>K}^{<A>}=0, 
$$
$$
{\cal R}_{<B><D>K}^{<A>}=(-)^{|K<D>|}{P}_{<B>K<D>}^{<A>},{\cal R}%
_{J<C><D>}^I=S_{J<C><D>}^I, 
$$
$$
{\cal R}_{<B><C><D>}^I=0,{\cal R}_{J<C><D>}^{<A>}=0,{\cal R}%
_{<B><C><D>}^{<A>}=S_{BCD}^A 
$$
( see formulas (10) and (14) in \cite{vjph2}).

The locally adapted components ${\cal R}_{<\alpha ><\beta >}={\cal R}ic(D)({%
\delta }_{<\alpha >},{\delta }_{<\beta >})$ (we point that in general on
dvs-bundles ${\cal R}_{<\alpha ><\beta >}\ne {(-)}^{\mid <\alpha ><\beta
>\mid }{\cal R}_{<\beta ><\alpha >})$ of the Ricci tensor are as follows: 
$$
{\cal R}_{IJ}=R_{IJK}^K,{\cal R}_{I<A>}=-{}^{(2)}{P_{I<A>}}=-{P}_{IK<A>}^K%
$$
$$
{\cal R}_{<A>I}={}^{(1)}{P_{<A>I}}={P}_{<A>I<B>}^{<B>},{\cal R}%
_{<A><B>}=S_{<A><B><C>}^{<C>}=S_{<A><B>}. 
$$
For the scalar curvature, ${\check {{\cal R}}}=Sc(D)=G^{<\alpha ><\beta >}%
{\cal R}_{<\alpha ><\beta >},$ we have 
$$
Sc(D)=R+S, 
$$
where $R=g^{IJ}R_{IJ}$ and $S=h^{<A><B>}S_{<A><B>}.$

The Einstein--Cartan equations on dvs-bundles are written as 
$$
{\cal R}_{<\alpha ><\beta >}-{\frac 12}G_{<\alpha ><\beta >}{\check {{\cal R}%
}}+{\lambda }G_{<\alpha ><\beta >}={\kappa }_1{\cal J}_{<\alpha ><\beta >},%
\eqno(35) 
$$
and 
$$
T_{<\beta ><\gamma >}^{<\alpha >}+{{G_{<\beta >}}^{<\alpha >}}{{T^{<\tau >}}%
_{<\gamma ><\tau >}}- 
$$
$$
{(-)}^{\mid <\beta ><\gamma >\mid }{{G_{<\gamma >}}^{<\alpha >}}{{T^{<\tau >}%
}_{<\beta ><\tau >}}={\kappa }_2{Q^{<\alpha >}}_{<\beta ><\gamma >},\eqno(36)
$$
where ${\cal J}_{<\alpha ><\beta >}$ and ${Q_{<\beta ><\gamma >}^{<\alpha >}}
$ are respectively components of energy-momentum and spin-density of matter
ds--tensors on la-space, ${\kappa }_1$ and ${\kappa }_2$ are the
corresponding interaction constants and ${\lambda }$ is the cosmological
constant. To write in an explicit form the mentioned matter sources of
la-supergravity in (35) and (36) there are necessary more detailed studies
of models of interaction of superfields on locally anisotropic superspaces
(in previous sections we presented details for a class of osculator
s--bundles; further generalizations with an explicit writing out of terms
higher order anisotropic interactions of s--fields on an arbitrary
dvs--bundle is connected with cumbersome calculations and formulas; we omit
such considerations in this paper).

Equations (35), can be split into base--- and fibre--components, 
$$
R_{IJ}-{\frac 12}(R+S-{\lambda })g_{IJ}={\kappa }_1{\cal J}%
_{IJ},{}^{(1)}P_{<A>I}={\kappa }_1{}^{(1)}{\cal J}_{<A>I},\eqno(37) 
$$
$$
S_{<A><B>}-{\frac 12}(S+R-{\lambda })g_{<A><B>}={{\kappa }_2}{\tilde {{\cal J%
}}}_{<A><B>},{}^{(2)}P_{I<A>}=-{{\kappa }_2}{}^{(2)}{{\cal J}_{I<A>}}, 
$$
are a supersymmetric higher order, with cosmological term, generalization of
the similar ones presented in \cite{ana86,ma87,ma94}, with prescribed
N-connection and h(hh)-- and v(vv)--torsions. We have added algebraic
equations (36) in order to close the system of s--gravitational field
equations (really we have also to take into account the system of
constraints for N--connection (see details and formula (19) in \cite{vjph2})
if locally anisotropic s--gravitational field is associated to a d--metric
(5)).

We point out that on la--superspaces the divergence $D_{<\alpha >}{\cal J}%
^{<\alpha >}$ does not vanish (this is a consequence of generalized Bianchi
and Ricci identities (15),(16) (17) and (18) from \cite{vjph2}). The
d-covariant derivations of the left and right parts of (35), equivalently of
(37), are as follows:%
$$
D_{<\alpha >}[{\cal R}_{<\beta >}^{\cdot <\alpha >}-{\frac 12}({\check {%
{\cal R}}}-2{\lambda }){\delta }_{<\beta >}^{\cdot <\alpha >}]= 
$$
$$
\left\{ 
\begin{array}{rl}
{\lbrack {{R_J}^I}-{\frac 12}{({R+S-2{\lambda })}}{{{\delta }_J}^I}]}_{\mid
I}+{}^{(1)}{P^{<A>}}_{I\perp <A>}=0, &  \\ 
{\lbrack {{S_{<B>}}^{<A>}}-{\frac 12}{({R+S-2{\lambda })}}{{{\delta }_{<B>}}%
^{<A>}}]}_{\perp <A>}-{}^{(2)}{P^I}_{<B>\mid I}=0, &  
\end{array}
\right. 
$$
where%
$$
{}^{(1)}{P^{<A>}}_J={}^{(1)}{P_{<B>J}}g^{<A><B>},{}^{(2)}{P^I}_{<B>}={}^{(2)}%
{P_{J<B>}}g^{IJ}, 
$$
$$
{R^I}_J={R_{KJ}}g^{IK},{S^{<A>}}_{<B>}={S_{<C><B>}}h^{<A><C>}, 
$$
and 
$$
{D_{<\alpha >}}{\cal J}_{\cdot <\beta >}^{<\alpha >}={\cal U}_{<\alpha >},%
\eqno(38) 
$$
where 
$$
D_{<\alpha >}{\cal J}_{\cdot <\beta >}^{<\alpha >}=\left\{ 
\begin{array}{rl}
{{\cal J}_{\cdot J\mid I}^I+{}^{(1)}{\cal J}_{\cdot J\perp <A>}^{<A>}}={%
\frac 1{{\kappa }_1}}{{\cal U}_J}, &  \\ 
{{}^{(2)}{\cal J}_{\cdot <A>\mid I}^I+{\cal J}_{\cdot <A>\perp <B>}^{<B>}}={%
\frac 1{{\kappa }_1}}{{\cal U}_{<A>}}, &  
\end{array}
\right. 
$$
and%
$$
{\cal U}_{<\alpha >}={\frac 12}(G^{<\beta ><\delta >}{{\cal R}_{<\delta
><\varphi ><\beta >}^{<\gamma >}}{{\cal T}_{\cdot <\alpha ><\gamma
>}^{<\varphi >}}-\eqno(39) 
$$
$$
{(-)}^{\mid <\alpha ><\beta >\mid }G^{<\beta ><\delta >}{{\cal R}_{<\delta
><\varphi ><\alpha >}^{<\gamma >}}{{\cal T}_{\cdot <\beta ><\gamma
>}^{<\varphi >}}+{{\cal R}_{\cdot <\varphi >}^{<\beta >}}{{\cal T}_{\cdot
<\beta ><\alpha >}^{<\varphi >}}). 
$$
So, it follows that ds-vector ${\cal U}_\alpha $ vanishes if d-connection
(34) is torsionless.

No wonder that conservation laws for values of energy--momentum type, being
a consequence of global automorphisms of spaces and s--spaces, or,
respectively, of theirs tangent spaces and s--spaces (for models on curved
spaces and s--spaces), on la--superspaces are more sophisticate because, in
general, such automorphisms do not exist for a generic local anisotropy. We
can construct a higher order model of supergravity, in a way similar to that
for the Einstein theory if instead an arbitrary metric d--connection the
generalized Christoffel symbols ${\tilde \Gamma }_{\cdot \beta \gamma
}^\alpha $ (constructed in a usual manner, but by using derivations (1)
instead of partial derivations, see explicit formulas (24) from \cite{vjph2}%
) are used. This is a locally anisotropic supersymmetric model on the base
s-manifold $\widetilde{M}$ which looks like locally isotropic on the total
space of a dvs--bundle. More general supergravitational models which are
locally anisotropic on the both base and total spaces can be generated by
using deformations of d-connections (see (25) in \cite{vjph2}). In this case
the vector ${\cal U}_\alpha $ from (39) can be interpreted as a
corresponding source of generic local anisotropy satisfying generalized
conservation laws of type (38).

More completely the problem of formulation of conservation laws for both
locally isotropic and anisotropic higher order supergravity can be solved in
the frame of the theory of nearly autoparallel maps of dvs-bundles (with
specific deformations of d-connections and in consequence of torsion and
curvature), which have to generalize our constructions from \cite{vcl96,vog}.

We end this section with the remark that field equations of type (35),
equivalently (37), for higher order supergravity can be similarly introduced
for the particular cases of higher order anisotropic s--spaces provided with
metric structure of type (5) with coefficients parametrized as for higher
order prolongations of the Lagrange, or Finsler, s--spaces (see details in 
\cite{vjph2}).

\section*{7. Gauge Like Locally Anisotropic Super\-gra\-vi\-ty}

The aim of this section is to introduce a set gauge like gravitational
equations (which are equivalent to Einstein equations on dvs--bundles (35)
if well defined conditions are satisfied). This model will be a higher order
anisotropic supersymmetric extension of our constructions for gauge
la-gravity \cite{vlasg,vg} and of affine--gauge interpretation of the
Einstein gravity \cite{p,pd,pon,ts,v87}.

The great part of theories of locally isotropic s-gravity are formulated as
gauge supersymmetric models based on supervielbein formalism (see \cite
{nie,sal,wes,west}). A similar model of supergravity on osculator s--bundles
has been considered in section 4. Here we shall analyzes a geometric
background for such theories on dvs--bundles. Let consider an arbitrary
adapted to N-connection frame $l_{<\underline{\alpha }>}(u)=(l_{\underline{I}%
}(u),l_{<\underline{C}>}(u))$ on ${\cal E}^{<z>}$ and s-vielbein matrix%
$$
l_{{<\alpha >}}^{<\underline{\alpha }>}=\left( 
\begin{array}{cccccc}
l_I^{\underline{I}} & 0 & ... & 0 & ... & 0 \\ 
0 & l_{A_1}^{\underline{A_1}} & ... & 0 & ... & 0 \\ 
... & ... & ... & 0 & ... & 0 \\ 
0 & 0 & 0 & l_{A_p}^{\underline{A_p}} & ... & 0 \\ 
0 & 0 & 0 & 0 & ... & 0 \\ 
0 & 0 & 0 & 0 & ... & l_{A_z}^{\underline{A_z}} 
\end{array}
\right) \subset GL_{n,k}^{<m,l>}(\Lambda )= 
$$
$$
GL(n,k,{\Lambda })\oplus GL(m_1,l_1,{\Lambda })\oplus ...\oplus GL(m_p,l_p,{%
\Lambda })\oplus ...\oplus GL(m_z,l_z,{\Lambda }) 
$$
for which 
$$
{\frac \delta {\delta u^{<\alpha >}}}=l{_{<\alpha >}}^{<\underline{\alpha }%
>}(u)l_{<\underline{\alpha }>}(u), 
$$
or, equivalently, ${\frac \delta {\partial x^I}}={l_I}^{\underline{I}}(u)l_{%
\underline{I}}(u)$ and ${\frac \delta {\partial y^{<C>}}}={l_{<C>}}^{<%
\underline{C}>}l_{<\underline{C}>}(u),$ and 
$$
G_{<\alpha ><\beta >}(u)=l{_{<\alpha >}}^{<\underline{\alpha }>}(u)l{%
_{<\beta >}}^{<\underline{\beta }>}(u){\eta }_{<{\underline{\alpha }><}{%
\underline{\beta }>}}, 
$$
where, for simplicity, ${\eta }_{<\underline{\alpha }><\underline{\beta }>}$
is a constant metric on vs-space $V^{n,k}\oplus V^{<l,m>}.$

By $LN({\cal E}^{<z>}{\cal )}$ is denoted the set of all adapted frames in
all points of sv-bundle ${\cal E}^{<z>}{\cal .}$ For a surjective s-map ${%
\pi }_L$ from $LN({\cal E}^{<z>}{\cal )}$ to ${\cal E}^{<z>}$ and treating $%
GL_{n,k}^{<m,l>}(\Lambda )$ as the structural s-group we define a principal
s--bundle, 
$$
{\cal L}N({\cal E}^{<z>})=(LN({\cal E}^{<z>}),{\pi }_L:LN({\cal E}^{<z>})\to 
{\cal E}^{<z>},GL_{n,k}^{<m,l>}({\Lambda })), 
$$
called as the s--bundle of linear adapted frames on ${\cal E}^{<z>}{\cal .}$

Let $I_{<\hat \alpha >}$ be the canonical basis of the sl-algebra ${\cal G}%
_{n,k}^{<m,l>}$ for a s-group\\ $GL_{n,k}^{<m,l>}({\Lambda })$ with a
cumulative index $<{\hat \alpha >}$. The structural coefficients\\ ${%
f_{<\hat \alpha ><\hat \beta >}}^{<\hat \gamma >}$ of ${\cal G}%
_{n,k}^{<m,l>} $ satisfy s-commutation rules 
$$
[I_{<\hat \alpha >},I_{<\hat \beta >}\}={f_{<\hat \alpha ><\hat \beta >}}%
^{<\hat \gamma >}I_{<\hat \gamma >}. 
$$
On ${\cal E}^{<z>}$ we consider the connection 1--form 
$$
{\Gamma }={{\Gamma }^{<\underline{\alpha }>}}_{<{\underline{\beta >}<}\gamma
>}(u)I_{<\underline{\alpha }>}^{<\underline{\beta }>}du^{<\gamma >},\eqno(40)
$$
where 
$$
{{\Gamma }^{<\underline{\alpha }>}}_{<\underline{\beta }><\gamma >}(u)=l{^{<%
\underline{\alpha }>}}_{<\alpha >}l{^{<\beta >}}_{<\underline{\beta }>}{{%
\Gamma }^{<\alpha >}}_{<\beta ><\gamma >}+l{^{<\underline{\alpha }>}}{\frac
\delta {\partial u^\gamma }}l{^{<\alpha >}}_{<\underline{\beta }>}(u), 
$$
${{\Gamma }^{<\alpha >}}_{<\beta ><\gamma >}{\quad }$ are the components of
the metric d--connection, s-matrix\\ $l{^{<\beta >}}_{<\underline{\beta }>}{%
\quad }$ is inverse to the s-vielbein matrix ${\quad }l{^{<\underline{\beta }%
>}}_{<\beta >},\quad $ and ${\quad }I_{<\underline{\beta >}}^{<\underline{%
\alpha }>}=\delta _{<\underline{\beta >}}^{<\underline{\alpha }>}$ is the
standard distinguished basis in SL--algebra ${\cal G}_{n,k}^{<m,l>}.$

The curvature ${\cal B}$ of the connection (40), 
$$
{\cal B}=d{\Gamma }+{\Gamma }\land {\Gamma }={\cal R}_{<\underline{\alpha }%
><\gamma ><\delta >}^{<\underline{\beta }>}I_{<\underline{\beta }>}^{<%
\underline{\alpha }>}{\delta u^{<\gamma >}}\land \delta u^{<\delta >} 
$$
has coefficients 
$$
{\cal R}_{<\underline{\alpha }><\gamma ><\delta >}^{<\underline{\beta }>}=l{%
^{<\alpha >}}_{<\underline{\alpha }>}(u)l{^{<\underline{\beta }>}}_{<\beta
>}(u){\cal R}_{<\alpha ><\gamma ><\delta >}^{<\beta >}, 
$$
where ${\cal R}_{<\alpha ><\gamma ><\delta >}^{<\beta >}$ are the components
of the ds--tensor.

In addition with ${\cal L}N({\cal E}^{<z>})$ we consider another s--bundle,
the bundle of adapted affine frames 
$$
{\cal A}N({\cal E}^{<z>})=(AN({\cal E}^{<z>}),{{\pi }_A}:AN({\cal E}%
^{<z>})\to {\cal E}^{<z>},{{AF}_{n,k}^{<m,l>}}({\Lambda })) 
$$
with the structural s--group ${AN}_{n,k}^{<m,l>}(\Lambda
)=GL_{n,k}^{<m,l>}(\Lambda )\odot {\quad }{\Lambda }^{n,k}\oplus {\Lambda }%
^{<m,l>}$ being a semidirect product (denoted by $\odot $ ) of $%
GL_{n,k}^{<m,l>}(\Lambda )$ and ${\Lambda }^{n,k}\oplus {\Lambda }^{<m,l>}.$
Because the LS--algebra ${{\cal A}f}_{n,k}^{<m,l>}$ of s--group $%
AF_{n,k}^{<m,l>}({\Lambda }),$ is a direct sum of ${\cal G}_{n,k}^{<m,l>}$
and ${\Lambda }^{n,k}\oplus {\Lambda }^{<m,l>}$ we can write forms on ${\cal %
A}N({\cal E}^{<z>})$ as $\Theta =({\Theta }_1,{\Theta }_2),$ where ${\Theta }%
_1$ is the ${\cal G}_{n,k}^{<m,l>}$--component and ${\Theta }_2$ is the $({%
\Lambda }^{n,k}\oplus {\Lambda }^{<m,l>})$--component of the form $\Theta .$
The connection (40) in ${\cal L}N({\cal E}^{<z>})$ induces a Cartan
connection $\overline{\Gamma }$ in ${\cal A}N({\cal E}^{<z>})$ (see the case
of usual affine frame bundles in \cite{bis,p,pd,pon} and generalizations for
locally anisotropic gauge gravity and supergravity in \cite{vg,vlasg} ).
This is the unique connection on dvs--bundle ${\cal A}N({\cal E}^{<z>})$
represented as $i^{*}{\overline{\Gamma }}=({\Gamma },{\chi }),$ where $\chi $
is the shifting form and $i:{\cal A}N({\cal E}^{<z>})\to {\cal L}N({\cal E}%
^{<z>})$ is the trivial reduction of dvs--bundles. If $l=(l_{<\underline{%
\alpha }>})$ is a local adapted frame in ${\cal L}N({\cal E}^{<z>})$ then ${%
\overline{l}}=i\circ l$ is a local section in ${\cal A}N({\cal E}^{<z>})$
and 
$$
{\overline{\Gamma }}=l\Gamma =(\Gamma ,\chi ),{\overline{{\cal B}}}={%
\overline{B}}{\cal B}=({\cal B},{\cal T}),\eqno(41) 
$$
where ${\chi }=e_{<\underline{\alpha }>}\otimes {l^{<\underline{\alpha }>}}%
_{<\alpha >}du^{<\alpha >},{\quad }e_{<\underline{\alpha }>}$ is the
standard basis in ${\Lambda }^{n,k}\oplus {\Lambda }^{<m,l>}$ and torsion $%
{\cal T}$ is introduced as 
$$
{\cal T}=d{\chi }+[{\Gamma }\land \chi \}={\cal T}_{\cdot <\beta ><\gamma
>}^{<\underline{\alpha }>}e_{<\underline{\alpha }>}du^{<\beta >}\land
du^{<\gamma >}, 
$$
${\cal T}_{\cdot <\beta ><\gamma >}^{<\underline{\alpha }>}={l^{<\underline{%
\alpha }>}}_{<\alpha >}{T^{<\alpha >}}_{\cdot <\beta ><\gamma >}$ are
defined by the components of the torsion ds--tensor.

By using a metric $G$ of type (5) on dvs--bundle ${\cal E}^{<z>}$ we can
define the dual (Hodge) operator ${*}_G:{\overline{\Lambda }}^{q,s}({\cal E}%
^{<z>})\to {\overline{\Lambda }}^{n-q,k-s}({\cal E}^{<z>})$ for forms with
values in LS--algebras on ${\cal E}^{<z>}$ (see details, for instance, in 
\cite{west}), where ${\overline{\Lambda }}^{q,s}({\cal E}^{<z>})$ denotes
the s--algebra of exterior (q,s)--forms on ${\cal E}^{<z>}{\cal .}$

Let operator ${*}_G^{-1}$ be the inverse to operator $*$ and ${\hat \delta }%
_G$ be the adjoint to the absolute derivation d (associated to the scalar
product for s--forms) specified for (r,s)--forms as 
$$
{\delta }_G={(-1)}^{r+s}{*}_G^{-1}\circ d\circ {*}_G. 
$$
Both introduced operators act in the space of LS--algebra--valued forms as 
$$
{*}_G(I_{<\hat \alpha >}\otimes {\phi }^{<\hat \alpha >})=I_{<\hat \alpha
>}\otimes ({*}_G{\phi }^{<\hat \alpha >}) 
$$
and 
$$
{\delta }_G(I_{<\hat \alpha >}\otimes {\phi }^{<\hat \alpha >})=I_{<\hat
\alpha >}\otimes {\delta }_G{\phi }^{<\hat \alpha >}. 
$$
If the supersymmetric variant of the Killing form for the structural
s--group of a s--bundle into consideration is degenerate as a s--matrix (for
instance, this holds for s--bundle ${\cal A}N({\cal E}^{<z>})$ ) we use an
auxiliary nondegenerate bilinear s--form in order to define formally a
metric structure ${G_{{\cal A}}}$ in the total space of the s--bundle. In
this case we can introduce operator ${\delta }_{{\cal E}}$ acting in the
total space and define operator ${\Delta }\doteq {\hat H}\circ {\delta }_{%
{\cal A}},$ where ${\hat H}$ is the operator of horizontal projection. After 
$\hat H$--projection we shall not have dependence on components of auxiliary
bilinear forms.

Methods of abstract geometric calculus, by using operators ${{*}_G},{{*}_{%
{\cal A}}},{{\delta }_G},{{\delta }_{{\cal A}}}$ and ${\Delta },$ are
illustrated, for instance, in \cite{p,pd} for locally isotropic spaces and
in \cite{vlasg,vg} for locally anisotropic, spaces. Because on superspaces
these operators act in a similar manner we omit tedious intermediate
calculations and present the final necessary results. For ${\Delta }{%
\overline{B}}$ one computers 
$$
{\Delta }{\overline{{\cal B}}}=({\Delta }{\cal B},{\cal {R\tau }}+{\cal R}%
i), 
$$
where ${\cal {R\tau }}={\delta }_G{\cal J}+{*}_G^{-1}[{\Gamma },{*}{\cal J}%
\} $ and 
$$
{\cal R}i={{*}_G^{-1}}[{\chi },{{*}_G}{\cal R}\}={(-1)}%
^{n+k+l_1+m_1+...+l_z+m_z}{{\cal R}_{<\alpha ><\mu >}}G^{<\alpha ><{\hat
\alpha >}}{e_{<\hat \alpha >}}{\delta u^{<\mu >}}.\eqno(42) 
$$
Form ${\cal R}i$ from (42) is locally constructed by using the components of
the Ricci ds--tensor (see Einstein equations (35) as one follows from the
decomposition with respect to a locally adapted basis ${\delta u^{<\alpha >}}
$ (2)).

Equations 
$$
{\Delta }{\overline{{\cal B}}}=0\eqno(43) 
$$
are equivalent to the geometric form of Yang--Mills equations for the
connection ${\overline{\Gamma }}$ (see (41)). D.A. Popov and L.I. Dikhin
proved \cite{p,pd} that such gauge equations coincide with the vacuum
Einstein equations if as components of connection form (41) the usual
Christoffel symbols are used. For spaces with local anisotropy the torsion
of a metric d--connection in general is not vanishing and we have to
introduce the source 1--form in the right part of (43) even gravitational
interactions with matter fields are not considered \cite{vg}.

Let us consider the locally anisotropic supersymmetric matter source ${%
\overline{{\cal J}}}$ constructed by using the same formulas as for ${\Delta 
}{\overline{{\cal B}}}$ when instead of ${\cal R}_{<\alpha ><\beta >}$ from
(42) is taken ${{\kappa }_1}({\cal J}_{<\alpha ><\beta >}-{\frac 12}%
G_{<\alpha ><\beta >}{\cal J})-{\lambda }(G_{<\alpha ><\beta >}-{\frac 12}%
G_{<\alpha ><\beta >}G_{<\tau >}^{\cdot <\tau >}).$ By straightforward
calculations we can verify that Yang--Mills equations 
$$
{\Delta }{\overline{{\cal B}}}=\overline{{\cal J}}\eqno(44) 
$$
for torsionless connection ${\overline{\Gamma }}=({\Gamma },{\chi })$ in
s-bundle ${\cal A}N({\cal E}^{<z>})$ are equivalent to Einstein equations
(35) on dvs--bundle ${\cal E}^{<z>}{\cal .}$ But such types of gauge like
la-super\-gra\-vi\-ta\-ti\-on\-al equations, completed with algebraic
equations for torsion and s--spin source, are not variational in the total
space of the s--bundle ${\cal A}L{({\cal E}}^{<z>}{)}.$ This is a
consequence of the mentioned degeneration of the Killing form for the affine
structural group \cite{bis,p,pd} which also holds for our la-supersymmetric
generalization. We point out that we have introduced equations (44) in a
''pure'' geometric manner by using operators ${*},{\quad }{\delta }$ and
horizontal projection ${\hat H}.$

We end this section by emphasizing that to construct a variational gauge
like supersymmetric la--gravitational model is possible, for instance, by
considering a minimal extension of the gauge s--group $AF_{n,k}^{m,l}({%
\Lambda })$ to the de Sitter s--group $S_{n,k}^{m,l}({\Lambda }%
)=SO_{n,k}^{m,l}({\Lambda }),$ acting on space ${\Lambda }_{n,k}^{m,l}\oplus 
{\cal R},$ and formulating a nonlinear version of de Sitter gauge s--gravity
(see \cite{ts,pon} for locally isotropic gauge gravity, \cite{vg} for a
locally anisotropic variant). 
\end{document}